%% file: main.tex
\documentclass[icol,sn-basic,Numbered]{sn-jnl}

\usepackage{float}
\usepackage{subcaption}
\usepackage{tikz}
\usepackage{svg}
\usepackage{amsmath,amssymb,amsfonts}
\usepackage{xcolor}
\usepackage{pgfplots}
\pgfplotsset{compat=1.18}
\usepackage{xcolor}
\usepackage{algorithm}
\usepackage{algpseudocode}
\usepackage{placeins}
\usepgfplotslibrary{groupplots}

\usetikzlibrary{calc,patterns,angles,quotes}
\graphicspath{{./}}

\newif\ifhighlight
\highlightfalse   

\newcommand{\myhighlight}[1]{%
  \ifhighlight
    \textcolor{purple}{#1}%
  \else
    #1%
  \fi
}

\newcommand{\myhighlightcolor}{%
  \ifhighlight
    \color{purple}%
  \fi
}

\begin{document}

\input{0_shortcuts.tex}
\input{0_abstract.tex}
\input{1_introduction.tex}
\input{2_mathematical_background.tex}

\input{3_application.tex}
\input{4_discussion.tex}
\appendix
\clearpage
\input{5_appendix.tex}

\section*{Acknowledgements}
The project was funded by the Bavarian State Ministry of Science and Arts within the \textit{SmartHeart} project.

\bibliography{literature}
\end{document}

%% file: 0_shortcuts.tex

\newcommand{\R}{\mathbb{R}}
\newcommand{\vv}{\boldsymbol{v}}

\newcommand{\vK}{\boldsymbol{K}}
\newcommand{\Kort}{K^{\perp}}
\newcommand{\vKort}{\boldsymbol{\boldsymbol{K}}^{\perp}}

\newcommand{\mom}{\alpha}
\newcommand{\vmom}{\boldsymbol{\mom}}
\newcommand{\dvmom}{\dot{\vmom}}
\newcommand{\vmomort}{\vmom^{\perp}}
\newcommand{\dvmomort}{\dot{\vmom}^{\perp}}

\newcommand{\cp}{c}
\newcommand{\vcp}{\boldsymbol{\cp}}
\newcommand{\ncp}{N_{cp}}

\newcommand{\vvort}{\dot{\vcp}^{\perp}}
\newcommand{\vvpar}{\dot{\vcp}^{\parallel}}

\newcommand{\vx}{\boldsymbol{x}}
\newcommand{\vdx}{\dot{\boldsymbol{x}}}
\newcommand{\vdxort}{\dot{\boldsymbol{x}}^{\perp}}
\newcommand{\vdxpar}{\dot{\boldsymbol{x}}^{\parallel}}

\newcommand{\trans}{\tau}
\newcommand{\vtrans}{\boldsymbol{\tau}}

\newcommand{\vrot}{\boldsymbol{r}}
\newcommand{\rotangle}{\omega}
\newcommand{\vangle}{\boldsymbol{\rotangle}}
\newcommand{\rotmat}{{\boldsymbol{R}}}
\newcommand{\xref}{\bar{x}}
\newcommand{\vxref}{\bar{\boldsymbol{x}}}

\newcommand{\eye}{\boldsymbol{I}}
\newcommand{\proj}{\boldsymbol{P}}
\newcommand{\port}{\proj^{\perp}}
\newcommand{\ppar}{\proj^{\parallel}}

\newcommand{\basis}{\boldsymbol{B}}
\newcommand{\vA}{ (\boldsymbol{B}^{T}\boldsymbol{B}) }

%% file: 0_abstract.tex
\title{\myhighlight{Surface matching for image-derived geometries with a projection-based separation of rigid body modes}}

\author[1,2]{\fnm{Ludwig}   \sur{Wagmüller\footnotemark[2] }\footnotetext[2]{These authors contributed equally to this work.} } \email{wagmueller.ludwig@hm.edu} 
\author[2]   {\fnm{Christian} \sur{Bilas\footnotemark[2] }}     \email{christian.bilas@tum.de}
\author[1]   {\fnm{Markus}    \sur{Gitterle}}  \email{markus.gitterle@hm.edu}
\author[1]   {\fnm{Michael}   \sur{Wibmer}}    \email{michael.wibmer@hm.edu}
\author*[2]  {\fnm{Michael}   \sur{Gee}}       \email{gee@tum.de}

\affil[1]{
    \orgdiv{Department of Mechanical, Automotive and Aeronautical Engineering},
    \orgname{Munich University of Applied Sciences},
    \orgaddress{\street{Lothstr. 64}, \city{Munich}, \postcode{80335}, \country{Germany}}
}
\affil[2]{
    \orgdiv{Mechanics \& High Performance Computing Group, School of Engineering and Design},
    \orgname{Technical University of Munich},
    \orgaddress{\street{Parkring 35}, \city{Garching}, \postcode{85748}, \country{Germany}}
}

\abstract{
    The large diffeomorphic deformation metric mapping (LDDMM) framework offers numerous techniques for analyzing shapes and images.
    Surface matching computes a non-rigid, smooth, and invertible mapping that registers one surface to another.
    The computed deformation may contain rigid body motion, which frequently does not represent meaningful shape differences but is due to external factors like imaging modalities, such as the object's positioning during image acquisition.
    We present a simple extension to the LDDMM framework's surface matching that uniquely decomposes the computed deformation into rigid body modes (RBM) and RBM-free motion by introducing a suitable projection.
    This enables a focused analysis of each component individually and additionally helps to preserve the surface discretization quality.
    \myhighlight{Three sets of examples illustrate our approach.
    A small-scale example demonstrates the proposed method's effects, followed by an example using a single pair of heart geometries, and finally an example involving a cohort of 98 heart shapes.}
}

\keywords{surface matching, image registration, subspace projection, rigid body motion, large diffeomorphic deformation metric mapping}
\maketitle

%% file: 1_introduction.tex
\section{Introduction}

Establishing spatial correspondence between medical images and the embedded anatomical geometries is critical in many medical applications, offering significant benefits for diagnosis, treatment planning, and monitoring.
The geometries are typically derived through manual segmentation and are represented by their surface.
Several methods have been proposed to perform the registration task.
Their categorization is extensive, and a complete account is not provided here but can be found in \cite{viergever_survey_2016}.
However, we outline the key ideas of two commonly used registration categories: landmark-based and landmark-free registration.

In the landmark-based approach, usually, a trained expert places corresponding landmarks in all images to track their position across a dimension, e.g., time.
The geometry change of interest is subsequently interpolated.
However, the precise placement of landmarks in anatomical images is challenging.
It requires expert knowledge and suffers from operator bias.
Additionally, pathological conditions can severely alter the anatomy, making the identification of consistent reference points difficult.
A pure landmark description of the geometry change cannot represent intricate local changes.

Landmark-free approaches describe the evolution of geometry in an integral way.
This captures the motion of the whole geometry by utilizing sophisticated surface representations.
A well-established method can be contained in the LDDMM framework and is called surface matching \cite{hutchison_surface_2005, reuter_deformetrica_2018}.
It defines a smooth and invertible mapping between segmented surfaces where no point-to-point correspondence is assumed or established.
This mapping is expressed with a spatial motion field.
The term LDDMM was first proposed in \cite{beg_computing_2005-1} and has since evolved into a widely adopted and extended framework.
\myhighlight{A comprehensive mathematical treatment of shapes and diffeomorphic matching within this framework is given in the monograph by Younes \cite{younes_shapes_2019}.}
A historically motivated literature overview is given in \cite{polzin_large_2018}.
LDDMM has a wide field of applications which ranges from palaeoanthropology \cite{braga_efficacy_2019}, to brain \cite{lin_asymmetry_2022, durrleman_morphometry_2014, stouffer_projective_2022,ceritoglu_multi-contrast_2009,vialard_diffeomorphic_2012-1} and heart \cite{guan_using_2024,guigui_parallel_2022} structures to name a few examples.
\myhighlight{Partial surface registration has been studied in \cite{hsieh_diffeomorphic_2021,antonsanti_partial_2021,ben_amor_resnet-lddmm_2022}.}
\cite{arguillere_registration_2016} simultaneously incorporates multiple objects and \cite{arguillere_diffeomorphic_2016} enforces inequality constraints within a deformation model, including affine alignment.
\cite{stouffer_projective_2022} reconstructed dense 3D anatomical coordinates from 2D measurements using projective LDDMM.

In the examples above and in general, RBMs may be of interest for certain applications, but often need to be separated from shape-changing deformations.
This is because RBMs may arise from factors like patient positioning (e.g., lying down, standing), breathing, or other external influences rather than intrinsic shape changes.
Consequently, isolating RBM motion from the total motion can significantly enhance the information value of subsequent statistical and computational analyses.
\myhighlight{Other approaches have addressed rigid motions within diffeomorphic registration, either by jointly optimizing rigid and non-rigid components \cite{glaunes_thesis_2005} or by introducing deformation models with localized rigid transformations, such as diffeomorphic modules \cite{gris_subriemannian_2018} or higher order momenta \cite{sommer_higherorder_2012}. These approaches improve the registration to better capture rigid motions by extending the deformation model. }

\myhighlight{
Instead of adding new terms to the deformation model, in this work, we propose an extension of the LDDMM framework that explicitly separates RBM motion from the shape-changing motion within the matching framework.}
We start with an overview of the essential aspects of LDDMM for surface matching to introduce necessary notation.
Subsequently, we briefly repeat subspace projections to eliminate rigid body components from a given motion field and integrate them into the LDDMM framework.
Finally, we present a small-scale example and an application using heart geometries.

%% file: 2_mathematical_background.tex
\section{Surface matching with separation of rigid body modes}

This section briefly discusses the basics of LDDMM surface matching, following the works in \cite{hutchison_surface_2005, durrleman_morphometry_2014, durrleman_statistical_2009}.

\subsection{Surface matching framework}
\begin{figure}[ht]
    \centering
    \includegraphics[width=0.4\linewidth]{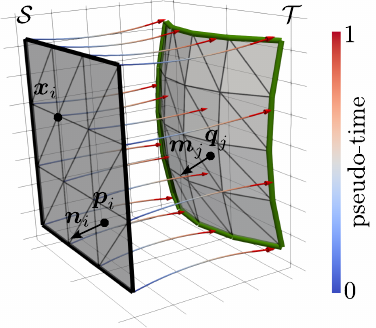}
    \caption{Surface Matching example. The source surface $\mathcal{S}$ depicted with a black contour is mapped onto the target surface $\mathcal{T}$ depicted with a green contour.
             The colored paths indicate the pseudo-temporal evolution of the source surface nodes $\vx_j$.
             Additionally, center points $\boldsymbol{p}_i$ and $\boldsymbol{q}_j$, along with normals $\boldsymbol{n}_i$ and $\boldsymbol{m}_j$, are displayed for $i$th and $j$th triangular element of the respective surface.}
    \label{fig:lddmm_intro}
\end{figure}

The objective is to identify a diffeomorphic transformation through pseudo-time $t \in [0,1]$ deforming a source surface $\mathcal{S}(t)$ to optimally represent a target surface $\mathcal{T}$.
Notably, this transformation is not limited to points located on the surfaces but instead distorts the entire ambient space and, thereby, the surfaces embedded in it.
In this work, we limit ourselves to directly summarizing the spatially discretized framework and skip going through the established underlying continuum theory for the sake of brevity.
Thus, a surface is represented by a mesh composed of nodes and triangular elements, as illustrated in Fig. \ref{fig:lddmm_intro}.

\subsubsection{Point motion through space and pseudo-time}

In LDDMM, the motion of an arbitrary point $\hat{\boldsymbol{x}}(t) \in \mathbb{R}^{3}$ is expressed as a weighted sum of radial basis functions $k_V$ located at control point positions $\vcp_k(t) \in \mathbb{R}^{3}$, where $k=1,2,\dots, \ncp$
\begin{equation}  \label{eq:arbitrary_motion}
    \dot{\hat{\boldsymbol{x}}}(t)=\sum_{k=1}^{\ncp} k_V(\hat{\boldsymbol{x}}(t),\vcp_{k}(t)) \vmom_{k}(t).
\end{equation}
$\vmom_{k}(t) \in \R^{3}$ are referred to as momenta and $N_{cp}$ is the number of control points.
Any point motion $\dot{\hat{\boldsymbol{x}}}(t)$ is an element of the pre-Hilbert space $V$.
In the case of infinite sums, $V$ can be completed into a reproducing kernel Hilbert space (RKHS); for more details, see, for example, \cite{hutchison_surface_2005, durrleman_morphometry_2014, glaunes_large_2008}.
With known initial positions, the point motion is parametrized only by $\vmom_{k}(t)$. For the sake of simplicity and readability, we omit the pseudo-time dependence in the following.
For the radial basis function $k_V$, we choose the Gaussian kernel with kernel width $\sigma_V \in \R$
\begin{equation} \label{eq:kernel_function}
    k_V(\hat{\boldsymbol{x}},\hat{\boldsymbol{x}}')
    = \exp\left(-\frac{\left\| \hat{\boldsymbol{x}}-\hat{\boldsymbol{x}}' \right\|_2^2}{\sigma^{2}_V} \right),
\end{equation}
where $\hat{\boldsymbol{x}}' \in \mathbb{R}^{3}$ denotes another arbitrary point position.

As control points are specific instances of arbitrary points $\hat{\boldsymbol{x}}$ in the ambient space, their trajectory adheres to the identical equation of motion \eqref{eq:arbitrary_motion} and reads
\begin{equation}
    \dot{\vcp}_{k}=\sum_{j=1}^{\ncp} k_V(\vcp_{k},\vcp_{j}) \vmom_{j}.
\end{equation}
Note that control points do not have to be located on the surface but can be distributed in its neighborhood.
One common approach for determining initial positions of control points is to distribute them uniformly across a grid within a bounding box of the source surface $\mathcal{S}$.

We are interested in the pseudo-temporal evolution of the source surface $\mathcal{S}$ defined by a total number of $N_x$ nodes, see Fig. \ref{fig:lddmm_intro}.
We denote the surface nodes by $\vx \in \R^{3N_x}$ and express their motion as
\begin{equation} \label{eq:point_motion}
    \dot{\vx}=\vK(\vx,\vcp) \vmom,
\end{equation}
where $\vcp, \vmom \in \R^{3\ncp}$ and $\vK(\vx,\vcp) \in \R^{3N_x \times 3\ncp}$ is called the kernel matrix.
It consists of point-wise main-diagonal blocks $\boldsymbol{K}_{ik}(\vx,\vcp) = \eye \: k_V(\vx_i,\vcp_k) \in \R^{3 \times 3}$, where $\eye \in \R^{3 \times 3}$ is the identity, $i=1,2,..N_x$ and $k=1,2,..\ncp$.
Similarly, using $\vK(\vcp,\vcp) \in \R^{3\ncp \times 3\ncp}$  the control point motion can be written as
\begin{equation} \label{eq:geo_cp}
    \dot{\vcp}=\vK(\vcp,\vcp) \vmom.
\end{equation}

\subsubsection{Uniqueness of transformation} \label{sec:uniqueness_of_motion}

The transformation of point positions defined by the equations of motion is determined by their initial positions and the pseudo-time varying momenta $\vmom$, which are non-unique.
A unique transformation is obtained by minimizing the motion's (pseudo-) kinetic energy $E$.
To compute the kinetic energy, a scalar product of two vector fields $\vv = \vK ( \cdot , \vcp )\boldsymbol{\alpha}$ and $\vv' = \vK ( \cdot , \vcp' ) \vmom'$, where $\{\vcp, \vcp'\}$ are control point positions and $\{\vmom, \vmom'\}$ are corresponding momenta, is defined as
\begin{equation} \label{eq:scalaprod}
    \langle \vv, \vv' \rangle_{V} := \vmom^{T} \vK(\vcp, \vcp') \boldsymbol{\vmom'}.
\end{equation}
The total kinetic energy of the control point motion is subsequently defined as the pseudo-time integral
\begin{equation} \label{eq:kinE}
    E := \int_{0}^{1} \langle \dot{\vcp}, \dot{\vcp} \rangle_{V} \:dt = \int_{0}^{1} \vmom^{T} \vK(\vcp,\vcp) \vmom \: dt.
\end{equation}
The authors in \cite{durrleman_morphometry_2014} show that in order to minimize the kinetic energy, the gradient of the kinetic energy with respect to the momenta $\vmom$ must vanish, which leads to
\begin{equation} \label{eq:geo_mom}
    \dot{\vmom}=-\frac{1}{2}\vmom^{T}\nabla_{\boldsymbol{c}}\vK(\vcp,\vcp) \vmom,
\end{equation}
where $\nabla_{\vcp}$ is the gradient with respect to the control point positions $\vcp$.

With equations \eqref{eq:point_motion}, \eqref{eq:geo_cp}, and \eqref{eq:geo_mom}, the transformation is uniquely characterized by the set of initial conditions  $\{\boldsymbol{x}_0, \: \vcp_0,\: \vmom_0\}$.

\subsubsection{Similarity between two surfaces}

Here, the metric to assess the similarity between source surface $\mathcal{S}$ and target surface $\mathcal{T}$ utilizes the varifold metric $A(\mathcal{S},\mathcal{T}) \in \R$ between surfaces
\begin{equation} \label{eq:surface_metric}
    \begin{aligned}
        A(\mathcal{S}, \mathcal{T})
        = \left(\right. \langle \mathcal{S}, \mathcal{S} \rangle_{W^{*}}
        +  &\langle \mathcal{T}, \mathcal{T} \rangle_{W^{*}} \\
        - 2&\langle \mathcal{S}, \mathcal{T} \rangle_{W^{*}} \left.\right).
    \end{aligned}
\end{equation}
$W$ is a RKHS with its dual space $W^{*}$, see \cite{durrleman_morphometry_2014} for details
\footnote{In the case $\mathcal{S}$ and $\mathcal{T}$ are composed of multiple disjoint subsurfaces, the similarity metric \eqref{eq:surface_metric} expands to $A(\mathcal{S}, \mathcal{T}) = \sum_{i}^{N_{obj}} \varepsilon_i A(\mathcal{S}_{i}, \mathcal{T}_i)$ where $N_{obj}$ is the number of subsurfaces each individually weighted by $\varepsilon_i$. For more details, the reader is referred to \cite{charon_varifold_2013, hutchison_surface_2005, durrleman_statistical_2009}.}.
The discrete inner product is given as
\begin{equation}
    \langle \mathcal{S}, \mathcal{T} \rangle_{W^{*}}
    = \sum\limits_i^{N_e^{\mathcal{S}}} \sum\limits_j^{N_e^{\mathcal{T}}} k_W(\boldsymbol{p}_i, \boldsymbol{q}_j) \frac{\left(\boldsymbol{n}^T_i \boldsymbol{m}_j \right)^2}{\left|\boldsymbol{n}_i \right| \left| \boldsymbol{m}_j \right|}.
\end{equation}
$\boldsymbol{p}_i$ and $\boldsymbol{q}_j \in \R^{3}$ denote the center points while $\boldsymbol{n}_i$ and $\boldsymbol{m}_j \in \R^{3}$ denote the normal vectors of each source $\mathcal{S}$ and target $\mathcal{T}$ triangular element, respectively.
The norm of $\boldsymbol{n}_i$ and $\boldsymbol{m}_j$ is half the area of the corresponding triangle.
See Fig. \ref{fig:lddmm_intro} for an example.
$N_e^{\mathcal{S}}$ and $N_e^{\mathcal{T}}$ are the number of elements.
Analogous to $k_V$, $k_W$ is a radial basis function.
We choose the Gaussian kernel with kernel width $\sigma_w \in \R$.
As we are interested in the similarity between the transformed source $\mathcal{S}$ and target $\mathcal{T}$, we assess \eqref{eq:surface_metric} at $t=1$ and write $A(\mathcal{S}, \mathcal{T})\Big|_{t=1}$.
The final optimization problem reads
\begin{equation} \label{eq:current_method}
    \begin{aligned}
        &\min_{\vmom_0} \: A(\mathcal{S}, \mathcal{T})\Big|_{t=1} + \int_{0}^{1} \vmom^{T} \vK(\vcp, \vcp) \vmom \: dt, \\
        & \text{s.t.} \:
        \begin{cases}
            \begin{alignedat}{2}
                &\dot{\vcp}           &&=\:\:\:\vK(\vcp,\vcp) \vmom, \\
                &\dot{\vmom}          &&=-\frac{1}{2} \vmom^{T} \nabla_{\vcp} \vK(\vcp, \vcp)  \vmom, \\
                &\dot{\boldsymbol{x}} &&=\:\:\: \vK(\boldsymbol{x}, \vcp) \vmom.
            \end{alignedat}
        \end{cases}
    \end{aligned}
\end{equation}
In the following, we will call this method the \textit{standard method}.

\subsection{Decomposing motion via projection}

For completeness and clarity, a short repetition of subspace projections is given in App.~\ref{app:subProj} based on \cite{banerjee_linear_2014}. We decompose the mesh node motion $\boldsymbol{\dot{x}} \in V$ into RBM motion $\boldsymbol{\dot{x}}^\parallel \in V^\parallel$ and RBM-free motion $\boldsymbol{\dot{x}}^\perp \in V^\perp$.
The corresponding projection matrix $\proj^\parallel$ and its orthogonal complement $\proj^\perp$ are computed from the basis $\basis$, see \cite{vanvek_convergence_2001}. $\basis$ represents RBMs and is defined as
\begin{equation} \label{eq:projector_basis_stack}
        \basis(\boldsymbol{x}) =\left[\begin{array}{c}
            \hat{\basis}(\boldsymbol{x}_1) \\
            \hat{\basis}(\boldsymbol{x}_2) \\
            \vdots \\
            \hat{\basis}(\boldsymbol{x}_{N_x})
        \end{array}\right],
\end{equation}
with $\hat{\basis}$ for each $\boldsymbol{x}_i = [x_i, y_i, z_i]^T$
\begin{equation} \label{eq:projector_basis}
        \hat{\basis}(\boldsymbol{x}_i) =\left[\begin{array}{cccccc}
            1 & 0 & 0 & 0 & z_i-\bar{z} & \bar{y}-y_i \\
            0 & 1 & 0 & \bar{z}-z_i & 0 & x_i-\bar{x} \\
            0 & 0 & 1 & y_i-\bar{y} & \bar{x}-x_i & 0 \\
        \end{array}\right].
\end{equation}
The six columns of $\hat{\basis}(\boldsymbol{x}_i) \in \R^{3 \times 6}$ are discrete representations of the six RBMs at $\boldsymbol{x}_i$.
The first three columns represent translations in $[x,y,z]$-directions and the last three represent linearized rotations around the Cartesian axis through a reference point $\vxref = [\bar x,\bar y,\bar z]^T$.

The projection matrix $\ppar$ is constructed as

\begin{equation} \label{eq:projector_definition}
    \begin{aligned}
        &\ppar(\boldsymbol{x}) =  \basis \left(\basis^T\basis\right)^{-1}\basis^T,\\
        &\proj^\perp(\boldsymbol{x}) =  \eye - \ppar(\boldsymbol{x}),
    \end{aligned}
\end{equation}
see e.g. \cite{banerjee_linear_2014,meyer_matrix_2000}.
This ensures that $\ppar$ projects any vector onto the six-dimensional subspace spanned by the RBM motion basis. The orthogonal complement $\proj^\perp$ captures any remaining motion components that cannot be expressed as a rigid body transformation.


\myhighlight{
\subsection{Surface matching with decomposed motion} \label{sec:first_mod}
We decompose the motion space $V$ into two orthogonal subspaces corresponding to RBM motions $V^\parallel$ and shape-changing motions $V^\perp$.
In the discrete setting, the infinite-dimensional RKHS $V$ of smooth vector fields on $\R^3$ is approximated via finite kernel sums centered at the time-evolving control points $\vcp$.
\newline
As the procedure is equivalent for $\dot{\vx} $ and $\dot{\vcp}$, we focus on the control point motion first, and omit the dependence on the $\vcp$ for brevity using $\vK = \vK(\vcp,\vcp), \port = \port(\vcp), \ppar = \ppar(\vcp)$. \\
We can use $\eye = \proj^\perp + \ppar$ from \eqref{eq:projector_definition} to expand the kernel such that 
\begin{equation}
\begin{aligned}
	\vK &= (\port+\ppar) \vK (\port+\ppar)\\
	&= \port \vK \port + \port \vK \ppar + \ppar \vK \port + \ppar \vK \ppar.
\end{aligned}
\end{equation}
As a modeling choice, the cross terms $\port \vK \ppar$ and $\ppar \vK \port$ are omitted, yielding a block-diagonal approximation of $\vK$, which allows for a decoupled analysis of the two subspaces:
\begin{equation} \label{eq:kernel_split}
    \begin{aligned}
	\tilde{\vK}
    &= \port \vK \port + \ppar \vK \ppar  \\
    &= \vKort + \vK^\parallel.
    \end{aligned}
\end{equation}
Appendix \ref{app:scalarproduct} shows that the subspaces $V^\perp, V^\parallel$ induced by  $\vKort, \vK^\parallel$ are finite dimensional approximations of a RKHS, too.\\
Following \eqref{eq:kernel_split} the decomposition of the control point motion $\dot{\vcp}$ from \eqref{eq:geo_cp} now reads
\begin{equation} \label{eq:motion_split}
    \begin{aligned}
        \dot{\vcp} 
        &:=  \vKort \vmom + \vK^\parallel \vmom \\
        &= \vvort + \vvpar.
    \end{aligned}
\end{equation}
We denote $\vvort$ as shape-changing motion and $\vvpar$ as RBM motion.
}

\myhighlight{
\subsubsection{Energy definition for the decomposed motion} 
In \eqref{eq:kinE}, the kinetic energy was defined with the help of the scalar product on $V$. 
Inserting the decomposition \eqref{eq:motion_split}, the scalar product now reads
\begin{equation}
    \begin{aligned}
        \langle \dot{\vcp}, \dot{\vcp} \rangle_V 
        &= \langle \vvort + \vvpar, \vvort + \vvpar \rangle \\
        &= \langle \dot{\vcp}, \dot{\vcp} \rangle_{V^\perp} 
        + \langle \dot{\vcp}, \dot{\vcp} \rangle_{V^\parallel} \\
        &= \vmom^T \vKort \vmom + \vmom^T \vK^\parallel \vmom,
    \end{aligned}
\end{equation}
With this, we can penalize the two components individually by weighting their contributions with factors $\gamma^\perp, \gamma^\parallel \in \R$, such that
\begin{equation} \label{eq:new_e_kin}
    \begin{aligned}
        \tilde{E} &=  \int_{0}^{1} \langle \dot{\vcp}, \dot{\vcp} \rangle_{V^\perp} 
        + \langle \dot{\vcp}, \dot{\vcp} \rangle_{V^\parallel} \: dt \\
        &= \gamma^\perp E^\perp + \gamma^\parallel E^\parallel
    \end{aligned}
\end{equation}
Analogously to subsection \ref{sec:uniqueness_of_motion}, in order for $\dot{\vmom}$ to minimize \eqref{eq:new_e_kin}, $\dot{\vmom}$ must fulfill
\begin{equation}
        \dot{\vmom} 
        = \dfrac{-1}{2(\gamma^\perp + \gamma^\parallel)} \vmom^T
        \left[ 
        \gamma^\perp \nabla_{\vcp} \vKort + \gamma^\parallel \nabla_{\vcp} \vK^\parallel
        \right] \vmom,
\end{equation}
as shown in App. \ref{app:mom_evolution}.
}


\subsection{Parametrization of RBMs} \label{sec:second_mod}
\begin{figure}[ht]
    \centering
    \begin{tikzpicture}[
        arrow style/.style={->,>=stealth, line width=1pt, shorten <=0.1cm, shorten >=0.1cm}
    ]

    \coordinate (A) at (-2,0);
    \coordinate (B) at (2,2);
    \coordinate (C) at (2.47,0.0);

    \node[above left] (Alabel) at (A) {$\vxref$};
    \node[above right] (Blabel) at (B) {$\vcp_k$};

    \draw[fill=black] (A) circle (0.1);
    \draw[fill=black] (B) circle (0.1);
    \draw[fill=black] (C) circle (0.1);

    \draw[arrow style] (A) -- (B) node[midway, above] {$\bar \vcp_k$};
    \draw[arrow style] (A) -- (C) node[midway, below] {$(\rotmat_{\vcp}(\vangle) \bar \vcp)_k$};

    \pic[draw, <-, >=stealth, "$\rotangle$", angle eccentricity=0.7, angle radius=40,line width=0.8pt] {angle = C--A--B};
    \pic[draw, arrow style,  <-, >=stealth, "$\vrot(\vangle{,} \vcp)_k$", angle eccentricity=1.16, angle radius=127] {angle = C--A--B};

    \end{tikzpicture}
    \caption{Exemplary visualization of the rotation of the $k^{th}$ control point $\vcp$ around a reference point $\vxref$.}
    \label{fig:rotfield}
\end{figure}
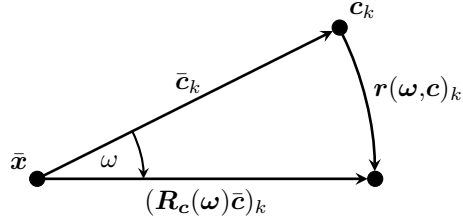

\myhighlight{
Having separated the RBM motion from the shape-changing motion, the RBM component can be analyzed independently. 
For such analyses, a compact and physically interpretable parametrization is essential, as it allows the magnitude and direction of RBM motion to be quantified directly.
We therefore represent $\vvpar$ as the sum of a translation $\vtrans_{\vcp}$ and a rotation $\vrot_{\vcp}$.
This yields a six–degree-of-freedom description consisting of three translation parameters $\hat \vtrans \in \R^{3}$ and three rotation parameters $\vangle \in \R^{3}$ that are constant in pseudo-time.
}
$\vvpar$ then reads

\begin{equation} \label{eq:rbm_motion}
    \vvpar := \vrot_{\vcp}(\vangle, \vcp) + \vtrans_{\vcp}(\hat \vtrans)
\end{equation}
The translation part $\vtrans_{\vcp}$ is also constant in space and reads
\begin{equation}
    \vtrans_{\vcp}(\hat \vtrans) =
    \begin{bmatrix}
        \boldsymbol{\boldsymbol{\hat \tau}} \\
        \boldsymbol{\boldsymbol{\hat \tau}} \\
        \vdots\\
        \boldsymbol{\boldsymbol{\hat \tau}} \\
    \end{bmatrix} \in \R^{3\ncp}.
\end{equation}
To ensure consistent dimensions, the three translation components $\boldsymbol{\hat \tau} = [\hat{\trans}_x, \hat{\trans}_y,\hat{\trans}_z ]^T$ are repeated for each control point.
The rotational component $\vrot_{\vcp} \in \R^{3\ncp}$ is parameterized by angles \( \vangle=[\rotangle_{x},\rotangle_{y},\rotangle_{z}]^{T} \) around the $x$,$y$ and $z$-axis in the reference point \( \vxref=[\bar{x},\bar{y},\bar{z}]^{T} \), and reads
\begin{equation} \label{eq:rotation_field}
    \vrot_{\vcp}(\vangle, \vcp) = \rotmat_{\vcp}(\vangle) \bar \vcp  - \bar \vcp,
\end{equation}
where $\bar \vcp = \vcp - \vxref$ are vectors rotated around the reference point $\vxref$ with rotation matrix $\rotmat_{\vcp}$, see Fig. \ref{fig:rotfield}.

\noindent
Analogous to the translation, the rotation matrix $\rotmat_{\vcp}\in \R^{3\ncp \times 3\ncp}$ consists of a point wise rotation matrix $\hat{\boldsymbol{R}} \in \R^{3 \times 3}$ repeated $\ncp$ times along its main diagonal, and thus is
\begin{equation}
    \rotmat_{\vcp}(\vangle) =
    \begin{bmatrix}
        \hat{\boldsymbol{R}}(\vangle) & \boldsymbol{0} & \cdots & \boldsymbol{0} \\
        \boldsymbol{0} & \hat{\boldsymbol{R}}(\vangle) & \cdots & \boldsymbol{0} \\
        \vdots & \vdots & \ddots & \vdots \\
        \boldsymbol{0} & \boldsymbol{0} & \cdots & \hat{\boldsymbol{R}}(\vangle)
    \end{bmatrix}.
\end{equation}
Since our motivation is to create an easily interpretable parametrization, we choose Euler angles to define $\hat{\boldsymbol{R}}(\vangle)=\hat{\boldsymbol{R}}_{x}(\rotangle_{x})\hat{\boldsymbol{R}}_{y}(\rotangle_{y})\hat{\boldsymbol{R}}_{z}(\rotangle_{z})$, consisting of three elemental rotations \( \hat{\boldsymbol{R}}_{i}(\rotangle_{i}), \: i \in \{x,y,z\} \) around the coordinate axes.
See App.~\ref{app:rotmats} for more details.
Note that while $\rotmat_{\vcp}(\vangle)$ is constant in space and pseudo-time,  $\vrot_{\vcp}(\vangle, \vcp)$ is not due to the spatial and pseudo-temporal dependence of $\vcp=\vcp(t)$. \\
\myhighlight{
Using the reparametrization \eqref{eq:rbm_motion} in the definition of $E^\parallel$ leads to 
\begin{align} \label{eq:kin_e_parallel_new}
    E^\parallel = \left\| \vrot_{\vcp}(\vangle, \vcp) + \vtrans_{\vcp}(\hat \vtrans) \right\|^2,
\end{align}
and the evolution of momenta (see App.~\ref{app:mom_evolution} for details) simplifies to
\begin{equation} \label{eq:geo_mom_new}
    \begin{aligned}
        \dot{\vmom} = 
        -\dfrac{1}{2\gamma} 
        \left( 
            \gamma^\perp \vmom^T \nabla_{\vcp} \vKort 
            + \gamma^\parallel \left( \rotmat_{\vcp}(\vangle) -\boldsymbol{I} \right) 
        \right) \vmom,
    \end{aligned}
\end{equation}
where $\gamma = \gamma^\perp + \gamma^\parallel$.
Note, while we employ Euler angles to reparametrize the RBM motion due to their simplicity, this step is optional, and the decomposition framework accommodates many other RBM parametrization possibilities.
}

\subsubsection{Final problem setup}
As mentioned above, we apply the same steps of subsection \ref{sec:first_mod} and \ref{sec:second_mod} to the motion of the mesh nodes $\boldsymbol{x}$ instead of control points $\vcp$, which results in
\begin{equation}
    \begin{aligned}
        \dot{\boldsymbol{x}}
        &= \dot{\boldsymbol{x}}^\perp + \dot{\boldsymbol{x}}^\parallel \\
        &:= \vKort(\boldsymbol{x}, \vcp) \vmom + \vrot_{\boldsymbol{x}}(\vangle, \boldsymbol{x}) + \vtrans_{\boldsymbol{x}}(\hat \vtrans).
    \end{aligned}
\end{equation}
Finally, the projection-based optimization problem reads

\begin{equation} \label{eq:my_method}
    \begin{aligned}
        &\min_{\vmom_0, \hat \vtrans_0, \vangle_0} 
        \: A(\mathcal{S}, \mathcal{T})\Big|_{t=1} 
        + \myhighlight{\gamma^\perp E^\perp + \gamma^\parallel E^\parallel} \\
        &\quad \text{s.t.} \:
        \begin{cases}
            \begin{alignedat}{2}
                &\dot{\vcp} &&=\:\:\:\vKort(\vcp,\vcp) \vmom + \vrot_{\vcp}(\vangle, \vcp) + \vtrans_{\vcp}(\hat \vtrans),\\
                &\dot{\vmom}  &&= -\dfrac{1}{2\gamma} 
        \left( 
            \gamma^\perp \vmom^T \nabla_{\vcp} \vKort 
            + \gamma^\parallel \left( \rotmat_{\vcp}(\vangle) -\boldsymbol{I} \right) 
        \right) \vmom, \\
                &\dot{\boldsymbol{x}} &&=\:\:\: \vKort(\boldsymbol{x}, \vcp) \vmom + \vrot_{\boldsymbol{x}}(\vangle, \boldsymbol{x}) + \vtrans_{\boldsymbol{x}}(\hat \vtrans).\\
            \end{alignedat}
        \end{cases}
    \end{aligned}
\end{equation}

To clarify, the variables $\boldsymbol{x}$, $\vmom$, $\vcp$, $\vrot$ evolve over pseudo-time $t$, while $\mathcal{T}$ and $\vtrans$ remain constant.
The similarity measure $A(\mathcal{S}, \mathcal{T})$ is evaluated at $t=1$ and the gradient $\nabla_{\vcp} \vKort$ is given in App.~\ref{app:kernel_gradient}.

%% file: 3_application.tex
\section{Examples}

First, we present a small-scale example featuring simple ellipsoid surfaces to demonstrate the impact of the decomposition.
\myhighlight{Next, we show an application using a single pair of heart geometries.
Finally, we present results on a cohort of 98 heart geometries.}
We solve all examples with the projection-based method \eqref{eq:my_method} and the standard method \eqref{eq:current_method} using an L-BFGS \cite{nocedal_updating_1980} optimizer.
We discretize the pseudo-time-dependent differential equations defining $\{\dot{\boldsymbol{x}}, \: \dot{\vcp},\: \dot{\vmom} \}$ with ten pseudo-time steps ($\Delta t = 0.1\text{ s}$) and solve them with  an explicit second-order Runge-Kutta \cite{runge_ueber_1895, burden_numerical_2016} scheme.
For the initial guess we set $\{\vmom_0, \hat \vtrans_0, \vangle_0\}$ to zero.
\myhighlight{To clarify the solution process, a pseudo-code for the projection-based method can be found in Alg.~\ref{alg:pseudo_code}}, see App.~\ref{app:pseudocode}.

\subsection{Ellipsoid example}
\begin{figure}[h]
    \centering
    \begin{subfigure}[c]{0.25\textwidth}
        \includegraphics[width=\linewidth]{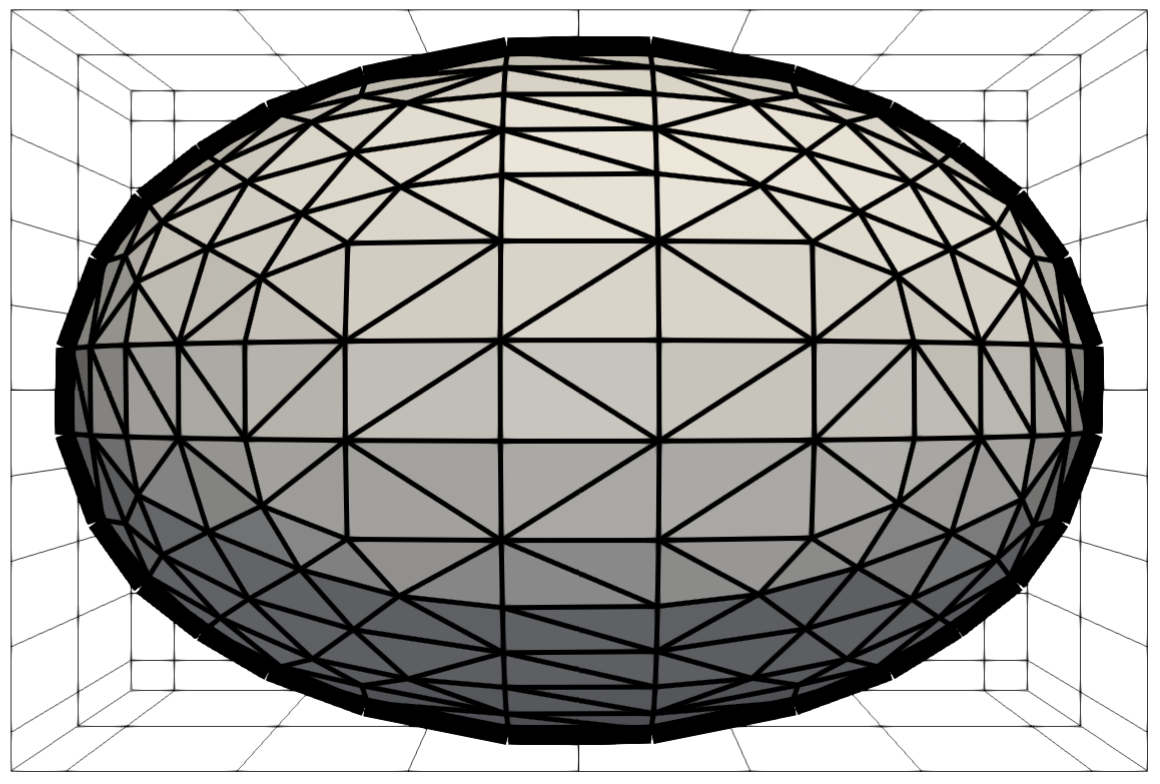}
        \caption{}
    \end{subfigure}
    \begin{subfigure}[c]{0.25\textwidth}
        \includegraphics[width=\linewidth]{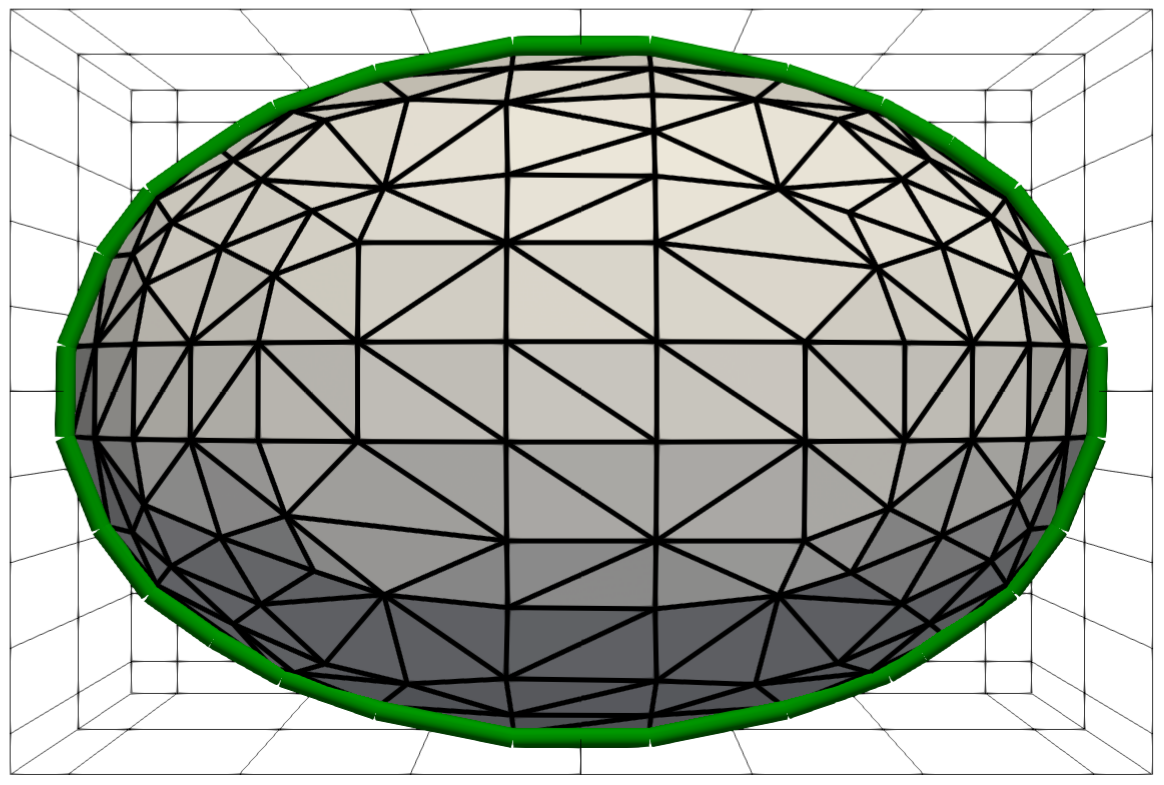}
        \caption{}
    \end{subfigure}
    \caption{Two ellipsoids discretized with triangular meshes, and their contour highlighted with thick colored lines. (a) is transformed with affine operations yielding three different source surfaces $\mathcal{S}_j$ depicted in Fig. \ref{fig:mini_example}. (b) is used as the target surface $\mathcal{T}$. }
    \label{fig:ellipsoids}
\end{figure}

\begin{figure*}[h]
    \centering
    \includegraphics[width=\textwidth]{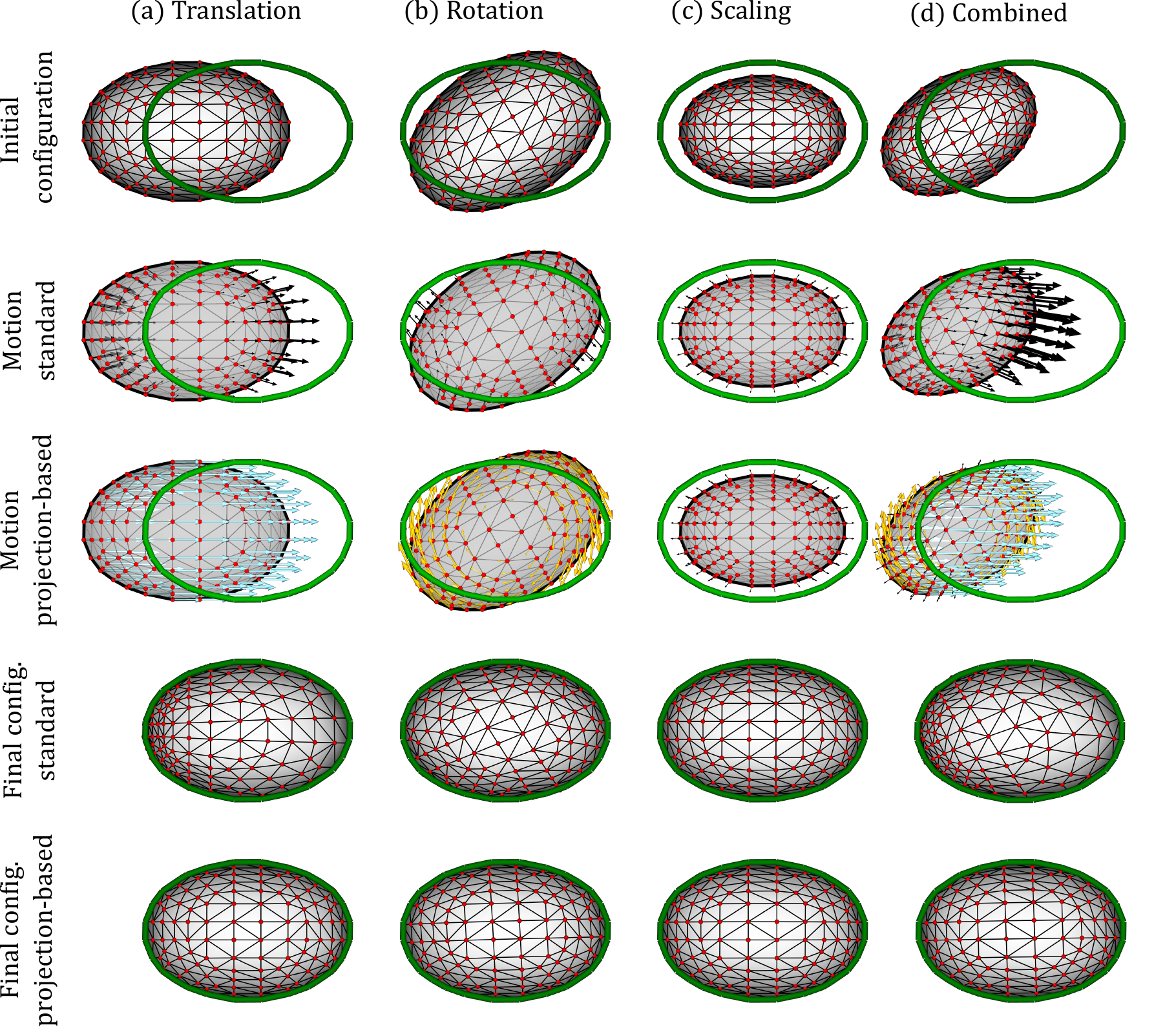}
    \caption{
        \myhighlight{
        Ellipsoid example.
        The first row shows the initial configuration of the source surfaces $\mathcal{S}_j(t=0)$ as a gray mesh and the targets as a green contour.
        The source surfaces are generated by four different affine operations.
        The second and third rows depict the motion of the standard and the projection-based method, respectively.
        Shape-changing motion is depicted in black, translation in blue, and rotation in orange.
        The arrows have been scaled to an observable range.
        The latter two rows show the resulting surfaces of each method at $t=1$.
        }
    }
    \label{fig:mini_example}
\end{figure*}

A three-dimensional ellipsoid with radii [$1350$ mm, $900$ mm, $900$ mm] is discretized with two different meshes and centered in the origin such that $\frac{1}{N_x}\sum_i \boldsymbol{x}_i = \boldsymbol{0}$.
We transform the surface in Fig. \ref{fig:ellipsoids}a with 
\myhighlight{four} affine operations to generate \myhighlight{four} initial source surfaces $\mathcal{S}_j$, see Fig. \ref{fig:mini_example}.
The transformations are:
\begin{itemize}
    \item Translation of $-400\text{ mm}$ in $x$ direction (Fig. \ref{fig:mini_example}a).
    \item Rotation of $-30^{\circ}$ around $y$-axis (Fig. \ref{fig:mini_example}b).
    \item $80\%$ scaling of all nodal coordinate vectors with respect to the ellipsoids center $\boldsymbol{x} \leftarrow 0.8\boldsymbol{x}$ (Fig. \ref{fig:mini_example}c).
    \item \myhighlight{A combination superimposing the above translation, rotation, and scaling.}
\end{itemize}
The initial source surfaces are subsequently matched onto the target in Fig. \ref{fig:ellipsoids}b.
Due to the smooth nature of the surfaces and the absence of local details, hyperparameters are set to $\sigma^{2}_W=\sigma^{2}_V=400\text{ mm}^2$, $\gamma^\perp = 10^{-2}, \gamma^\parallel = 10^{-4}$.
All surface nodes are designated as control points, with the reference point $\vxref$ coinciding with the center of the source ellipsoids $\mathcal{S}_j$, serving as the coordinate system's origin.

Given that translation and rotation constitute RBM transformations, the projection-based approach is anticipated to reconstruct these transformations without the component $\dot{\boldsymbol{x}}^{\perp}$.
Conversely, scaling represents a uniform distortion, expected to result in a motion without the component $\dot{\vx}^\parallel$.

Both methods successfully deform the source surfaces $\mathcal{S}_j(t=1)$ to accurately match the target surface $\mathcal{T}$, but the nodal distribution differs between the two approaches, see Fig. \ref{fig:mini_example}.
In the translation example (Fig. \ref{fig:mini_example}a), the standard method introduces distortion within the surface.
This distortion arises from the non-uniform motion depicted in the second row of Fig. \ref{fig:mini_example}.
In contrast, the projection-based method exhibits a spatially constant translation.
It is illustrated with blue arrows in the third row of Fig. \ref{fig:mini_example}.
Similarly, the projection-based method exclusively applies rotation.
Both methods, standard and projection-based, lead to an identical result for the scaling example.
In the following figures, we use black arrows for shape-changing motion, blue for translation, and orange for rotation.
The average nodal motion components are listed in Table~\ref{tab:ellipsoid_avg_motions}.
\begin{table}[h]
    \caption{
        Average motion ellipsoids [$\frac{\text{ mm}}{\Delta t}$].
        \label{tab:ellipsoid_avg_motions}
    }
    \myhighlightcolor
    \begin{tabular}{lcc}
        \hline
        Example  & Standard & Projection-based \\
         &method & method \\
        \hline
        \hline
        (a) Translation & $(\dot{\vcp})_{avg}=211$ & $(\vtrans)_{avg}=400$ \\
        (b) Rotation    & $(\dot{\vcp})_{avg}=68\phantom{0}$ & $(\vrot)_{avg}=240$ \\
        (c) Scaling     & $(\dot{\vcp})_{avg}=57\phantom{0}$ & $(\dot{\vcp})_{avg}^{\perp}=57\phantom{0}$ \\
        (d) Combined    & $(\dot{\vcp})_{avg}=233$ & $(\dot{\vcp})_{avg}^{\perp}=55\phantom{0}$\\
                    &                          & $(\vtrans)_{avg}=399$ \\
                    &                          & $(\vrot)_{avg}=233$ \\
        \hline
    \end{tabular}
\end{table}

\begin{figure*}
    \centering
    \includegraphics[width=0.65\textwidth]{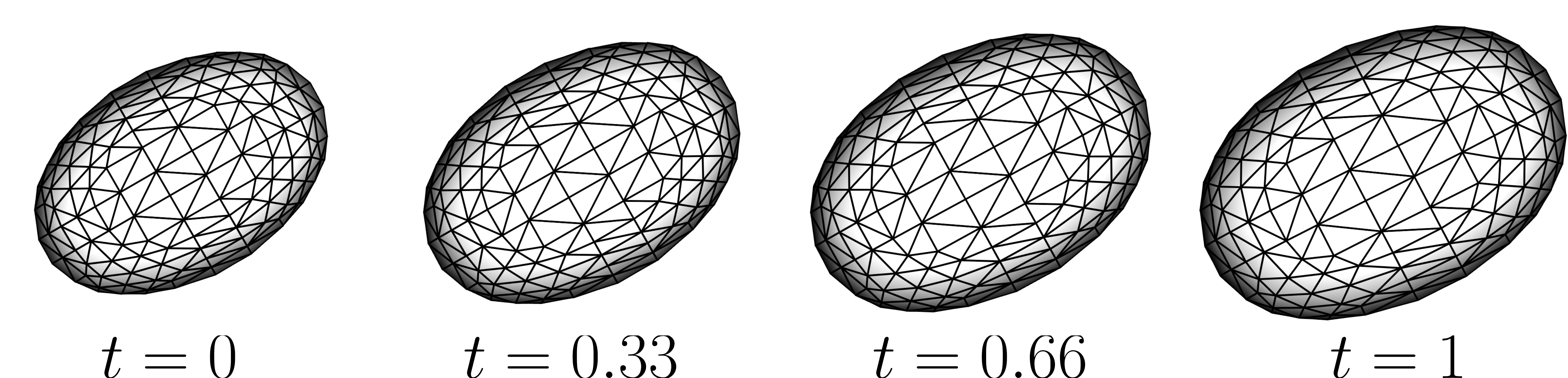}
    \caption{\myhighlight{Evolution of pure shape-changing motion over pseudo-time for the combined ellipsoid example (last column of Fig.~\ref{fig:mini_example}). As the projection-based method separates RBM from shape-changing motion, a focused analysis of the pure shape-changing motion is possible.}}
    \label{fig:mini_example_shape_change}
\end{figure*}

\myhighlight{As the projection-based method separates the RBM motion from the shape-changing motion, it is easy to remove the RBM component and focus on the pure shape-changing motion. Fig. \ref{fig:mini_example_shape_change} shows the pure shape-changing motion of the combined example.}

\begin{figure}[h]
    \centering
    \begin{subfigure}[c]{0.45\textwidth}
        \centering
        \includegraphics[width=\linewidth]{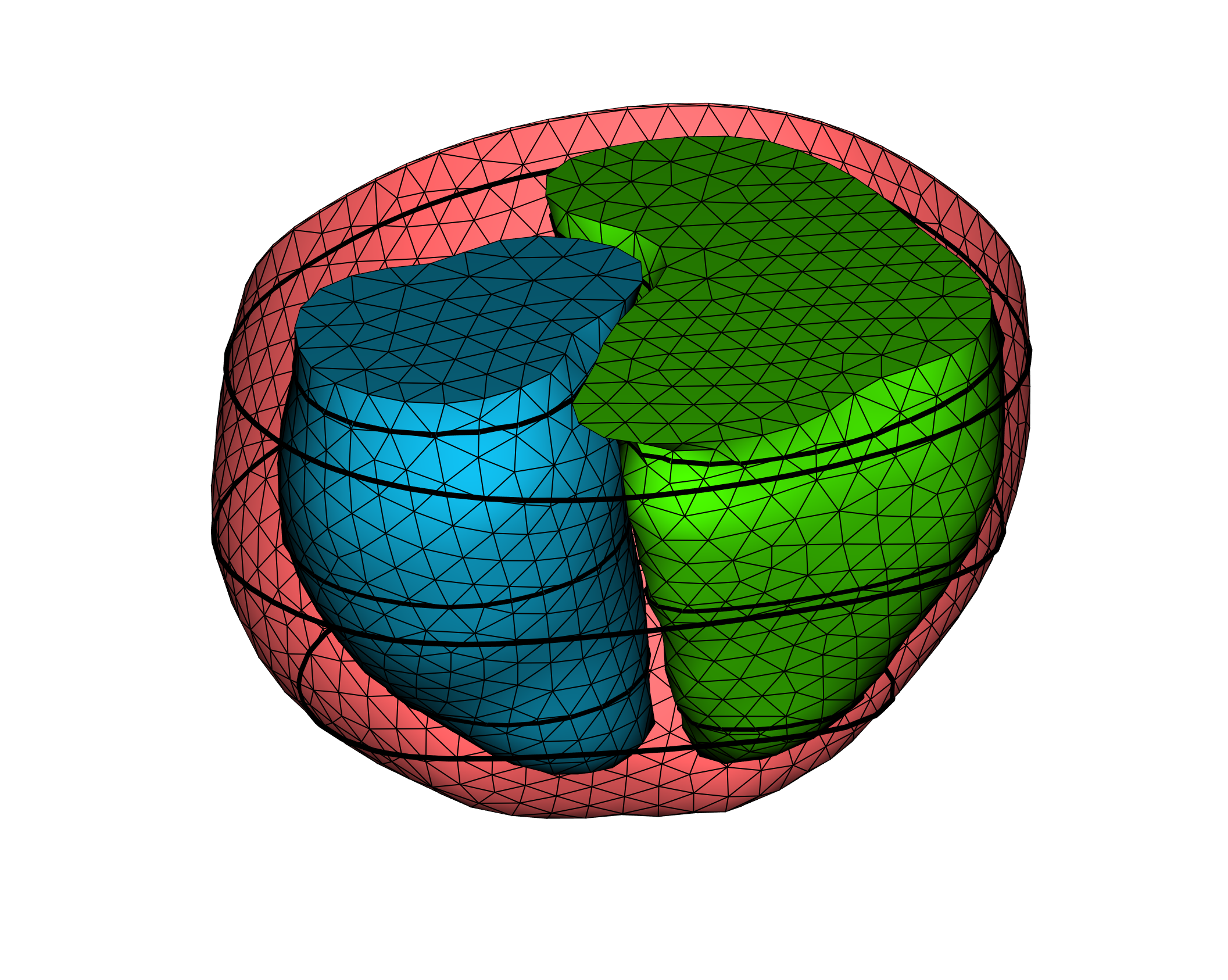}
        \caption{}
    \end{subfigure}%
    \begin{subfigure}[c]{0.45\textwidth}
        \centering
        \includegraphics[width=\linewidth]{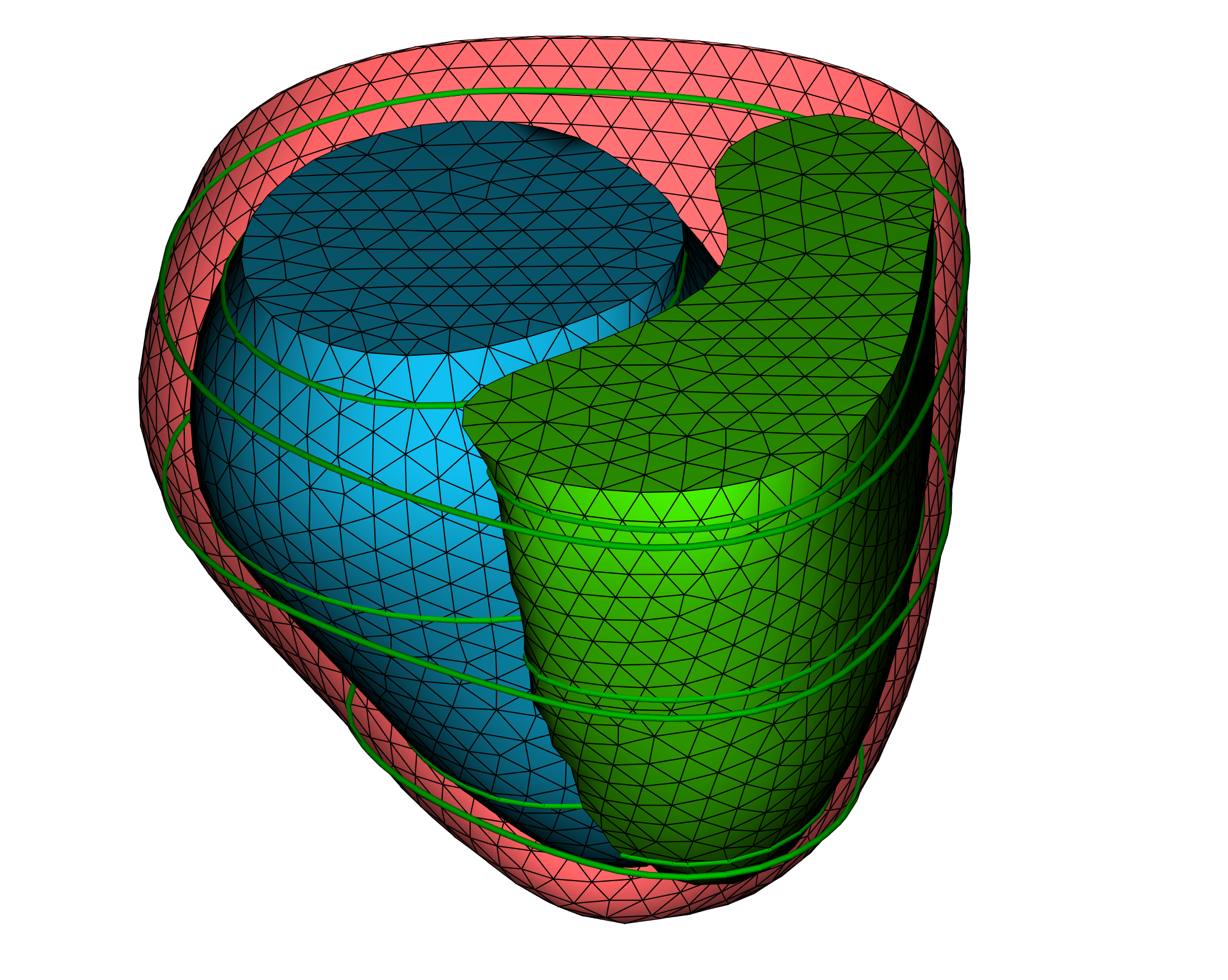}
        \caption{}
    \end{subfigure}
    \caption{Source (a) and target (b) heart geometry.
    Both consist of an epicardial (red), a left ventricular (blue), and a right ventricular (green) luminal subsurface.
    Control points are arranged as a uniform grid enclosing the source heart and are depicted as red spheres.
    \label{fig:heart_initial}}
\end{figure}

\FloatBarrier
\myhighlight{
\subsection{Single pair of heart geometries}} \label{sec:single_hearts}
When analyzing organ shapes over time, variations in patient positioning during CT or MRI introduce rigid body motion, which does not reflect actual anatomical changes.
This can obscure meaningful structural differences, especially when comparing scans of the same organ, such as the heart, at different time points.
For applications like growth analysis, it is essential to eliminate RBMs to ensure accurate alignment.
Therefore, we demonstrate the applicability of the proposed method to an example utilizing heart geometries obtained from two different points in time, see Fig. \ref{fig:heart_initial}.

\begin{figure*}[!htbp]
    \centering
    \begin{subfigure}{0.64\textwidth}
        \centering
        \includegraphics[width=\textwidth]{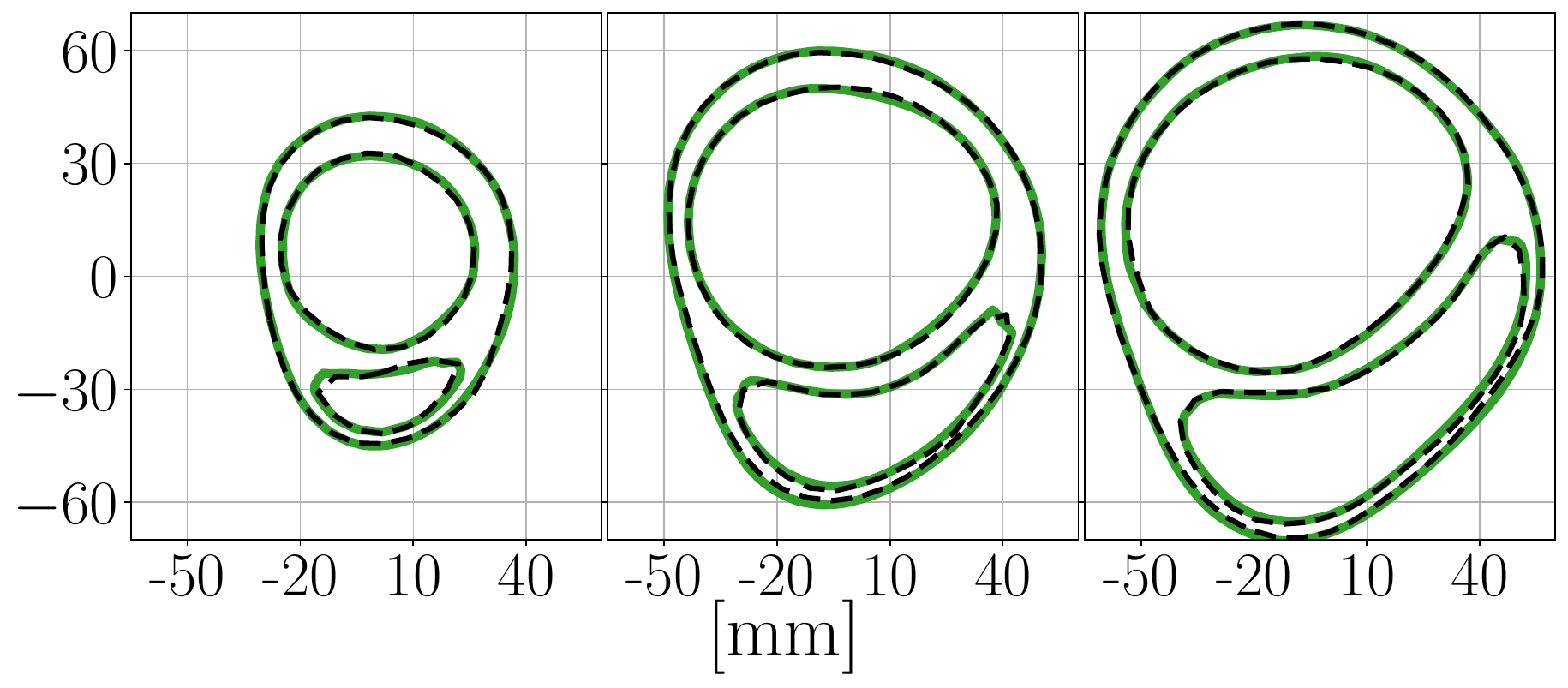}
        \caption{\myhighlight{Registration standard method}}
    \end{subfigure}
    \begin{subfigure}{0.35\textwidth}
        \centering
        \includegraphics[width=0.9\textwidth]{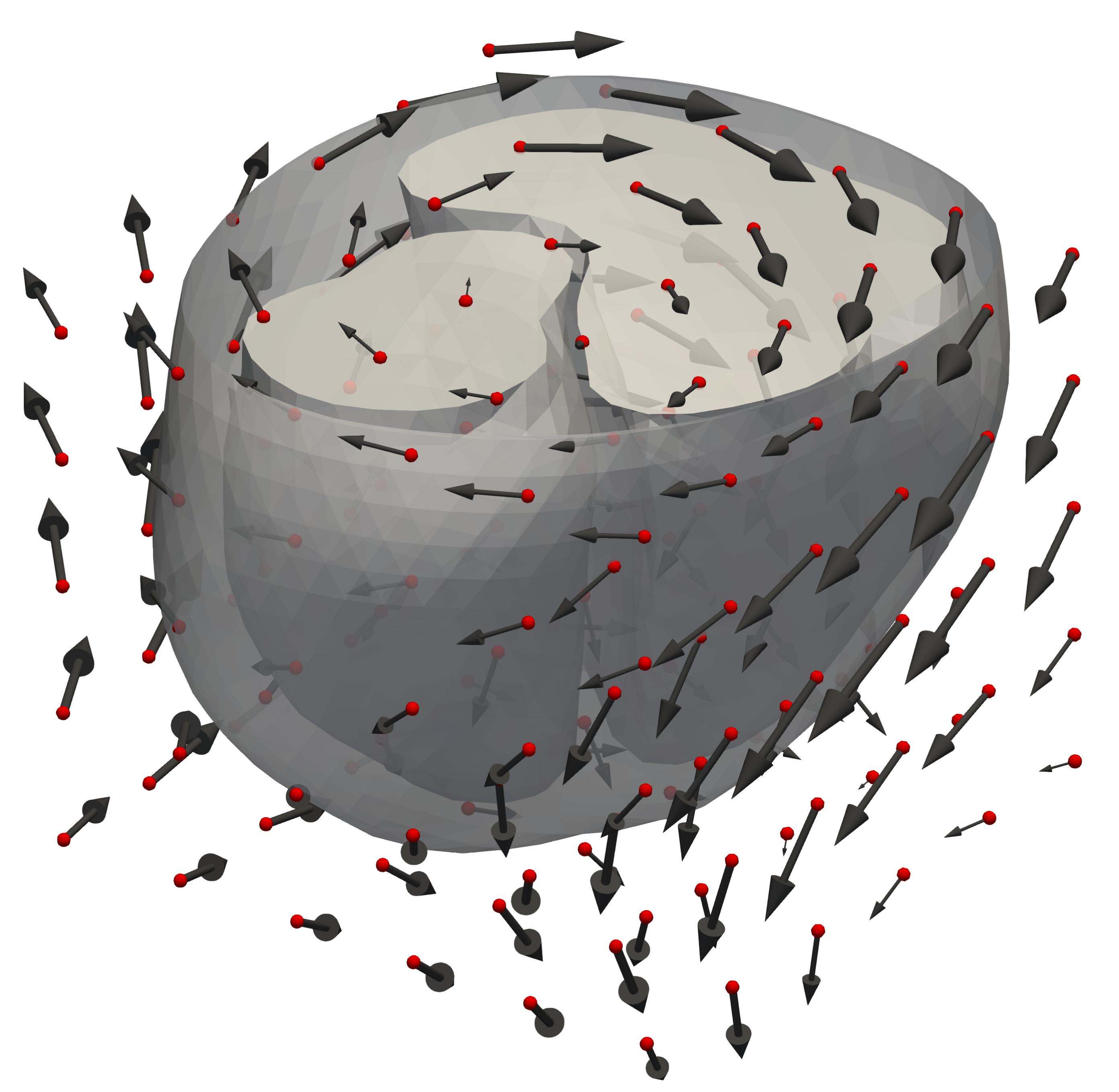}
        \caption{Motion standard method}
    \end{subfigure}

    \begin{subfigure}{0.64\textwidth}
        \centering
        \includegraphics[width=\textwidth]{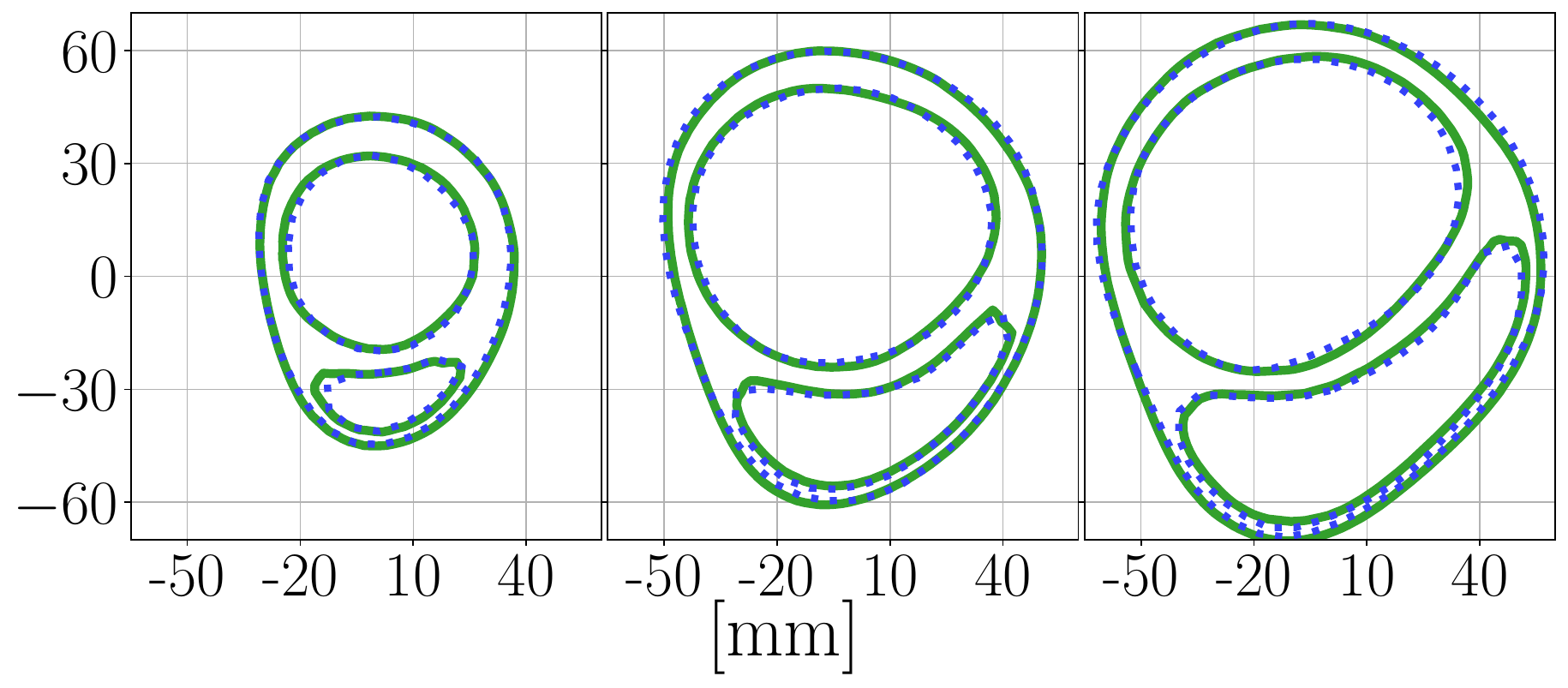}
        \caption{\myhighlight{Registration projection-based method}}
    \end{subfigure}
    \begin{subfigure}{0.35\textwidth}
        \centering
        \includegraphics[width=0.9\textwidth]{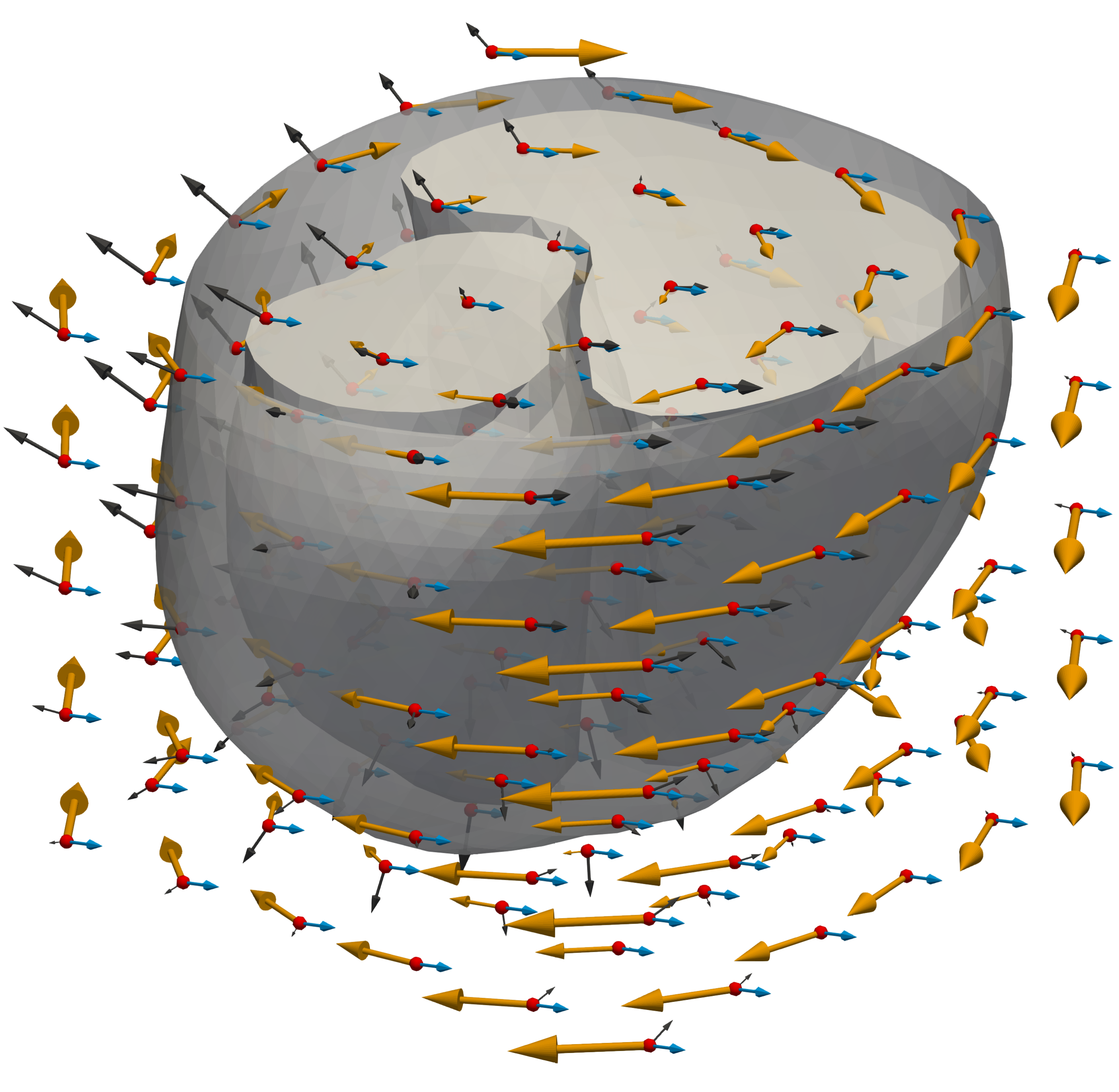}
        \caption{Motion projection-based method}
    \end{subfigure}

    \begin{subfigure}[c]{0.4\textwidth}
        \includegraphics[width=0.8\textwidth]{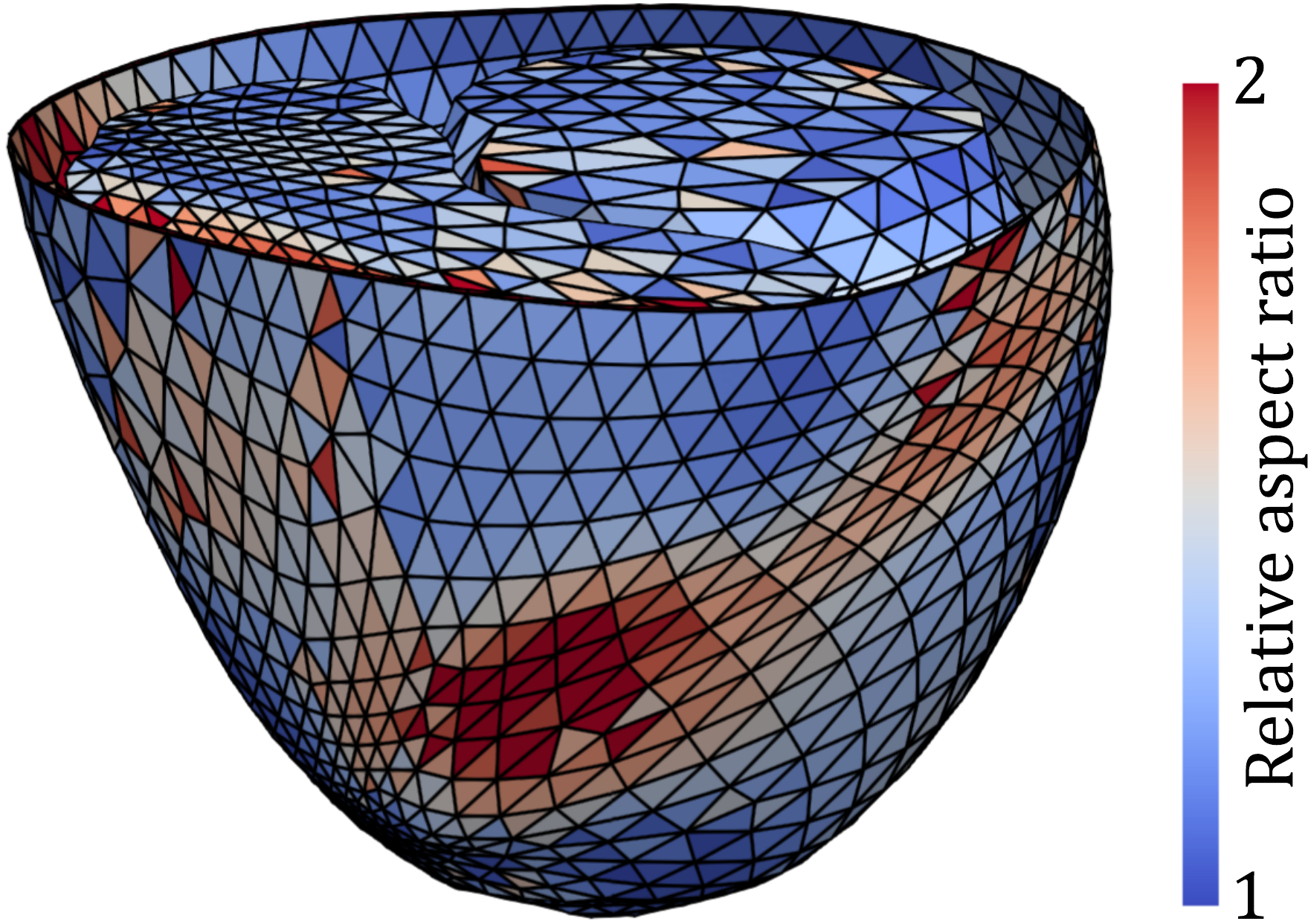}
        \caption{Mesh quality standard method}
    \end{subfigure}
    \hspace{1cm}
    \begin{subfigure}[c]{0.4\textwidth}
        \includegraphics[width=0.8\textwidth]{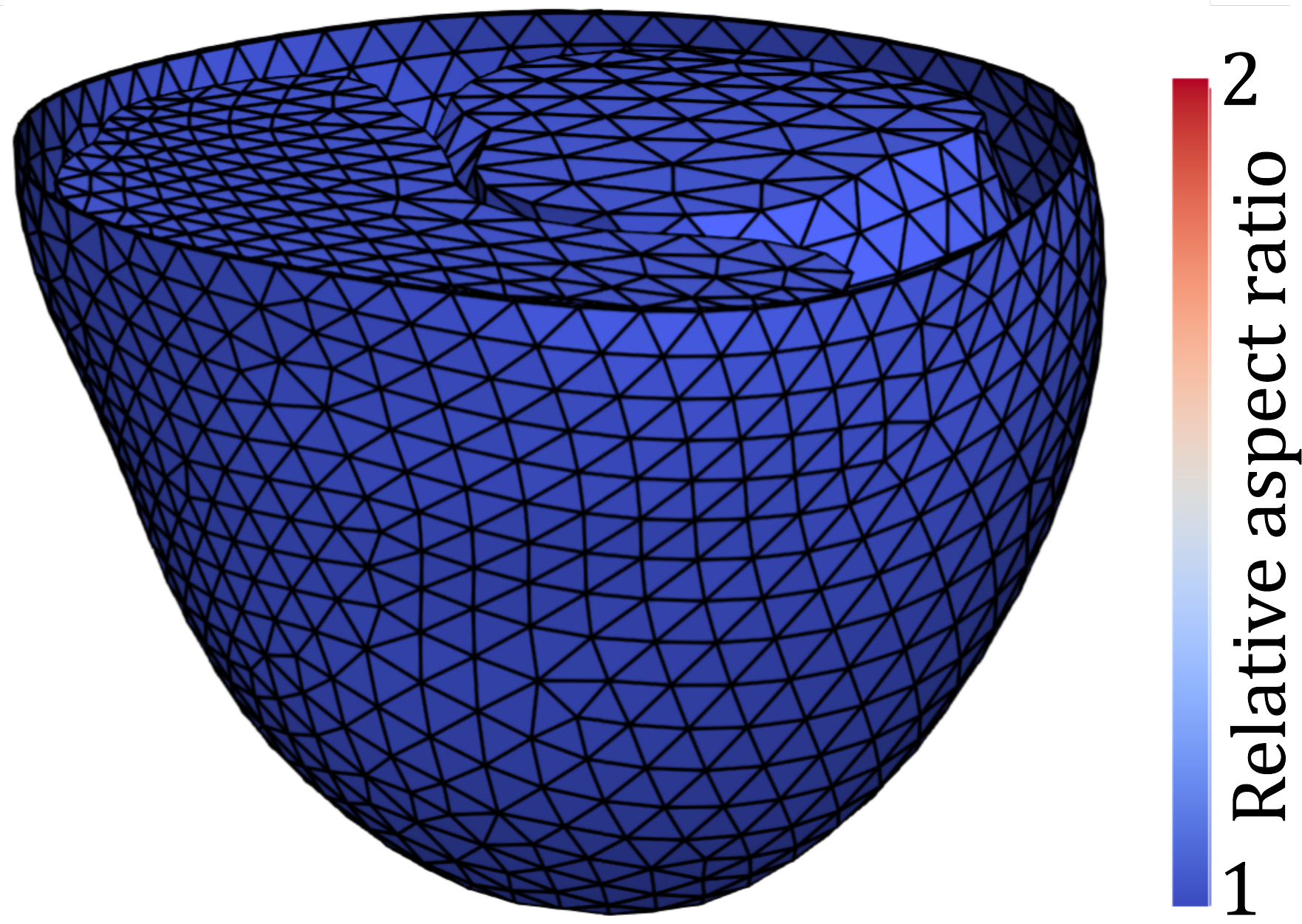}
        \caption{Mesh quality projection-based method}
    \end{subfigure}

    \caption{
    Heart example results.
    (a) and (c) illustrate the registration quality of both methods. The target heart is sliced at three uniformly distributed heights as depicted in Fig.~\ref{fig:heart_initial}b.
    Green depicts the target, and the dashed and dotted lines depict the deformed source surfaces at $t=1$ of the standard and projection-based method, respectively.
    (b) and (d) show the motion components at $t=0$.
    shape-changing motion is depicted in black, translation in blue, and rotation in orange.
    The arrows have been scaled to an observable range.
    (e) and (f) illustrate the mesh element deterioration due to the deformation.
    The surface triangles are colored with the relative aspect ratio.
    Red corresponds to a higher and blue to a lower mesh distortion.
    }
    \label{fig:heart_results}
\end{figure*}

The two geometries consist of the epicardial, left, and right ventricular luminal surfaces each.
All surfaces are discretized with an average element edge length of $6\text{ mm}$ and were manually segmented from CT images with a resolution of $1.2\text{ mm}$. The surfaces are centered at the coordinate origin such that $\frac{1}{N_x}\sum_i \boldsymbol{x}_i = \boldsymbol{0}$.
We choose the kernel width to approximately three times the average edge length $\sigma^{2}_W=\sigma^{2}_V=20\text{ mm}^2$ and $\gamma^\perp = 10^{-1}, \gamma^\parallel = 10^{-4}$.
The control points are arranged as a uniform grid with a spacing of $\sigma^{2}_W$ enclosing the source heart.
Fig. \ref{fig:heart_results} demonstrates that both the standard and projection-based method yield highly similar surface registrations accurately representing the target heart anatomy.
To quantify the registration quality, we define the node-to-node distance of the $i$th source node $\vx_i$ at $t=1$ as the minimal euclidean distance to each target node $\vx_j^{\mathcal{T}}$ for $j=1,..,N_x^{\mathcal{T}}$
\begin{equation}
    d_i=\min_j\left\|\vx_i - \vx_j^{\mathcal{T}}\right\|,
\end{equation}
where $N_x^{\mathcal{T}}$ is the number of target mesh nodes.

Table \ref{tab:solution_table} reveals that both methods achieve good matching quality.
The projection-based method exhibits a slightly larger error, but the difference is significantly below the spatial discretization and the underlyingCT image resolution.

\begin{table}[h]
    \caption{Node-to-node distance [mm]. \label{tab:solution_table}}
    \centering
    \begin{tabular}{lcc}
        \hline
        Node-to-node &  Standard  & Projection-based  \\
        distance $d_i$ & method & method \\
        \hline
        \hline
        Average & 0.5 & 0.7 \\
        Maximum & 4.1 & 4.4 \\
        \hline
    \end{tabular}
\end{table}

We use the aspect ratio $s$ of each triangular element to quantify its distortion at a given pseudo-time $t$.
It is defined as the scaled ratio between the element's smallest height $h_{min}$ and the element's longest edge length $l_{max}$
\begin{equation*}
    s(t)=\frac{\sqrt{3}}{2} \frac{h_{min}}{l_{max}}.
\end{equation*}
The bottom row of Fig.~\ref{fig:heart_results} visualizes that, similar to the ellipsoid example, the standard method internally distorts the surface more, by coloring the elements with their relative aspect ratio
\begin{equation} \label{eq:aspectratio}
    s_{rel}=\frac{s(t=1)}{s(t=0)}.
\end{equation}

Fig. \ref{fig:heart_results}d and illustrates the control point motion fields, composed of translation $\vtrans$, rotation $\vrot$, and RBM-free motion $\vvort$.
The dominant component is the rotational field, see Table~\ref{tab:heart_avg_motions}.
Only the RBM-free motion $\vvort$, depicted in black, distorts the surface.
The projection-based method significantly reduces the influence of this component and thus preserves the element quality better.

\begin{table}[h]
    \caption{
        Average motion hearts [$\frac{\text{ mm}}{\Delta t}$].
        \label{tab:heart_avg_motions}
    }
    \begin{tabular}{lcc}
        \hline
        Motion & Standard & Projection-based \\
        component & method & method \\
        \hline
        \hline
        Shape-changing & $(\dot{\vcp})_{avg}=34.0$ &  $(\dot{\vcp})_{avg}^{\perp}=11.2$ \\
        Translation &- & $(\vtrans)_{avg}=7.5\phantom{0}$\\
        Rotation    &- & $(\vrot)_{avg}=39.1$ \\
        \hline
    \end{tabular}
\end{table}

\begin{table}[h]
\myhighlightcolor
  \caption{\myhighlight{Extracted RBMs.}}
  \centering
  \begin{tabular}{l@{\hspace{16pt}}ccc@{\hspace{18pt}}cccc}
      \hline
      Component & Unit &x&y&z\\
      \hline
      \hline
      Translation & [mm] &  -0.2 &  -6.0 &  2.4   \\
      Rotation & [°] & -0.1 &  3.1 & 38.1 \\
      \hline
  \end{tabular}
  \label{tab:rbm_heart_pair}
\end{table}

\begin{figure*}[h]
    \centering
    \includegraphics[width=0.9\textwidth]{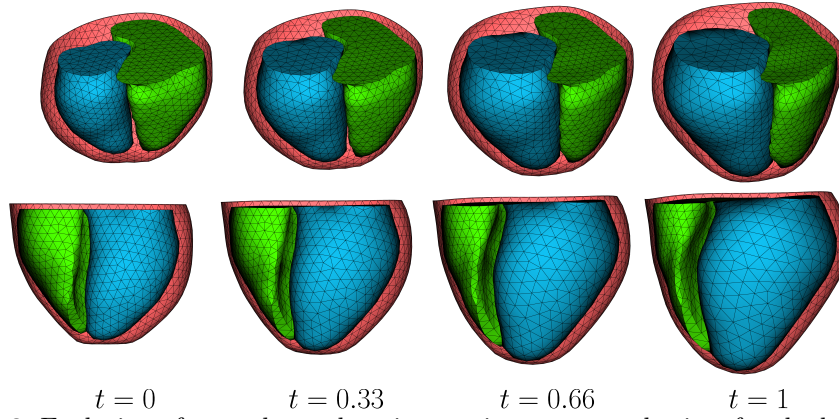}
    \caption{\myhighlight{Evolution of pure shape-changing motion over pseudo-time for the heart pair example.
    The top row shows the temporally resolved evolution in an angled view, and the bottom row depicts it in frontal view.}}
    \label{fig:heart_results_pure_shape}
\end{figure*}

\myhighlight{The separated RBM components are listed in Table~\ref{tab:rbm_heart_pair} and the effect of the remaining pure shape-changing motion is visualized in Fig.~\ref{fig:heart_results_pure_shape}.}

\FloatBarrier


\myhighlight{
\subsection{Heart cohort example}}

\begin{figure*}[h]
    \centering
    \begin{tikzpicture}
        \node[anchor=south west, inner sep=0] (svg) at (0,0) {\includegraphics[width=0.24\textwidth]{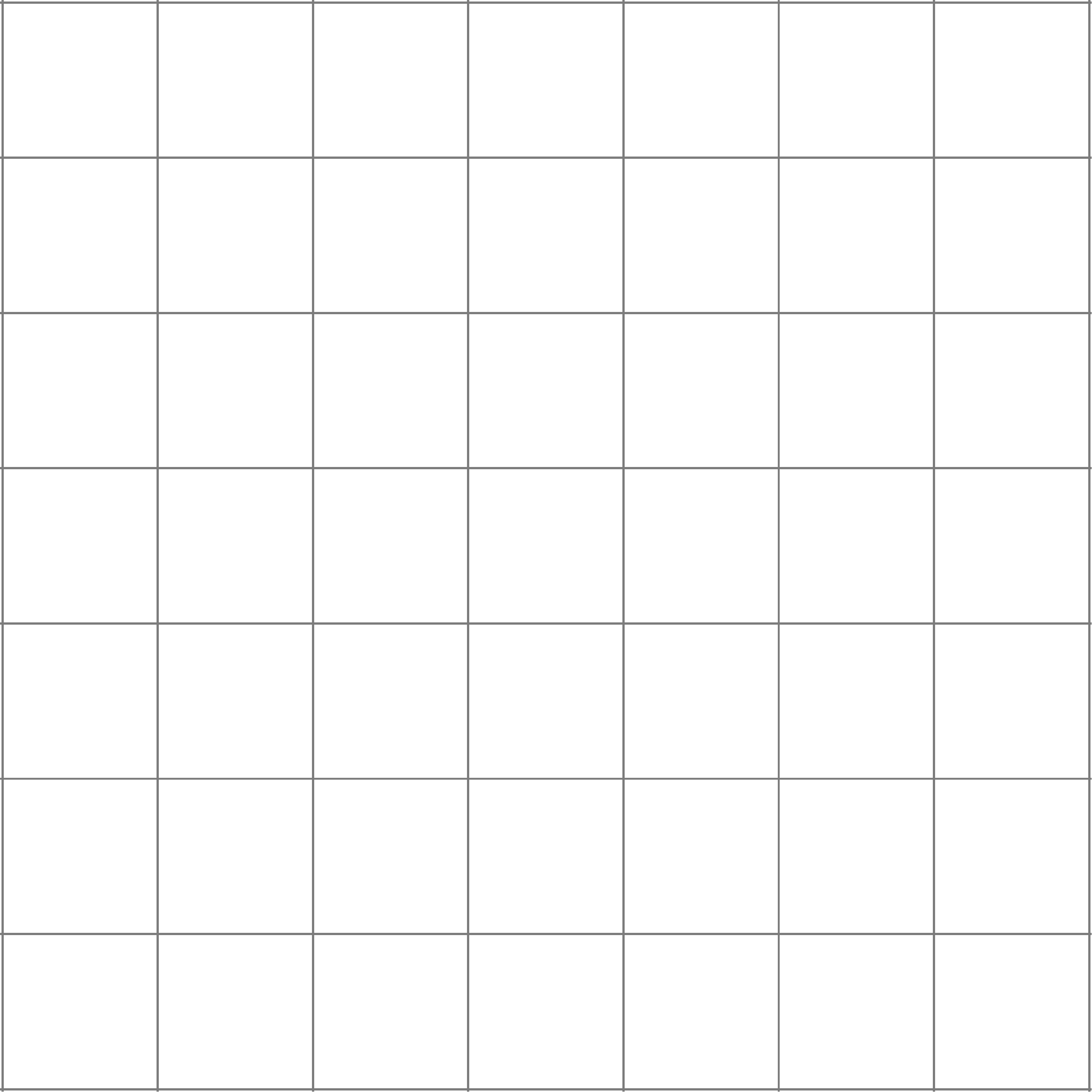}};
        \node[anchor=south west, inner sep=0] at (svg.south west) {\includegraphics[width=0.24\textwidth]{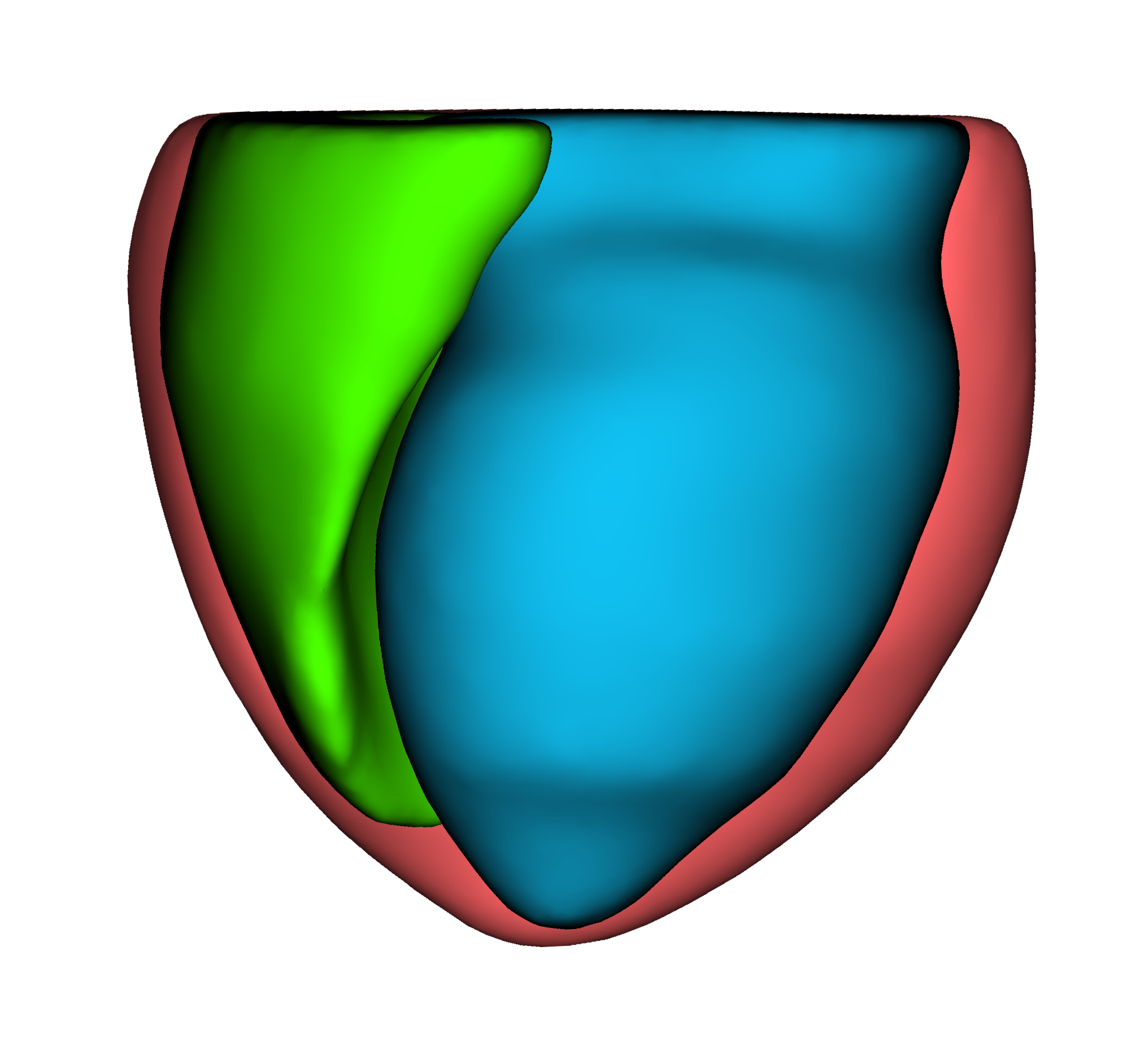}};
    \end{tikzpicture}
    \begin{tikzpicture}
        \node[anchor=south west, inner sep=0] (svg) at (0,0) {\includegraphics[width=0.24\textwidth]{grid.pdf}};
        \node[anchor=south west, inner sep=0] at (svg.south west) {\includegraphics[width=0.24\textwidth]{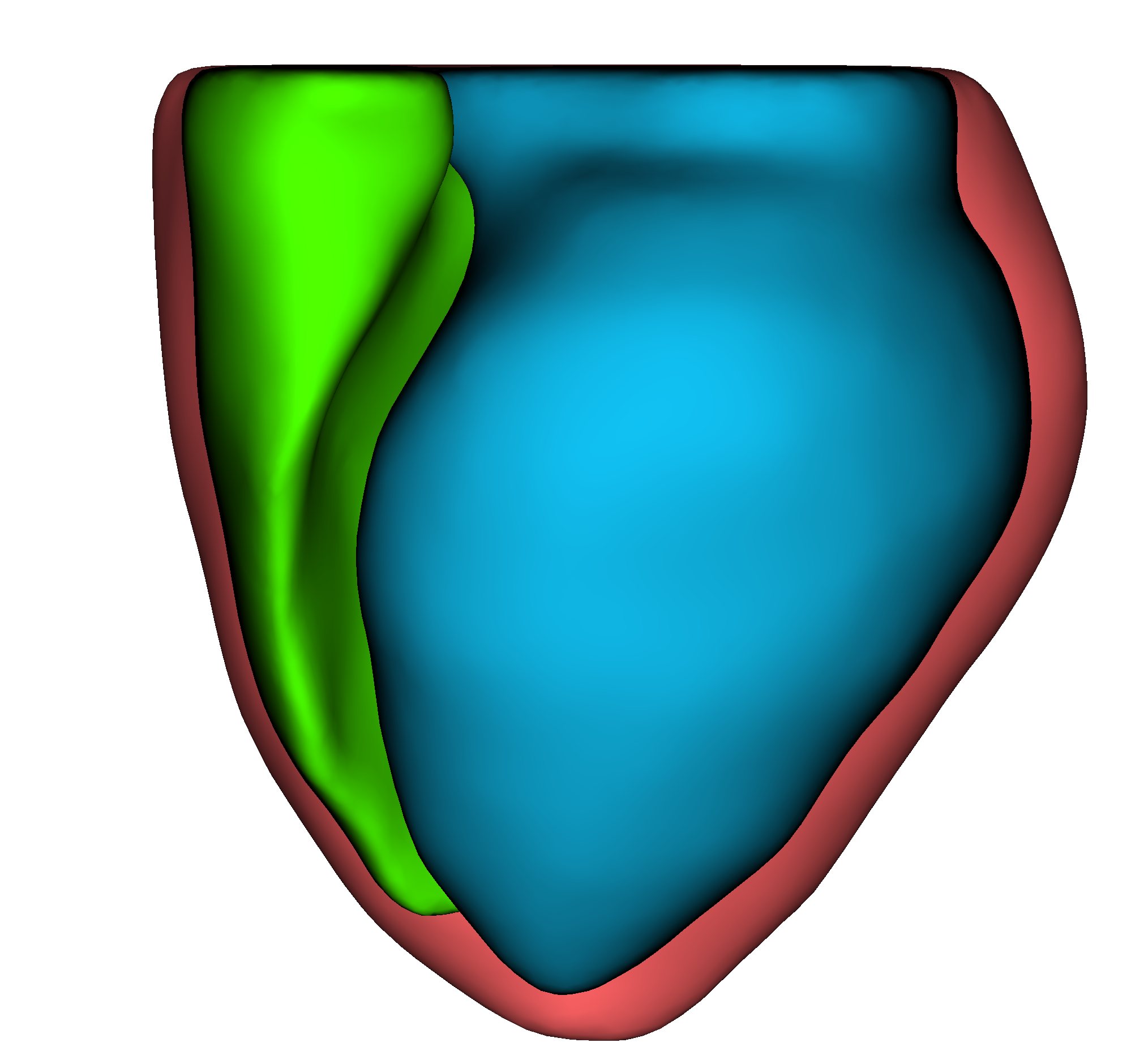}};
    \end{tikzpicture}
    \begin{tikzpicture}
        \node[anchor=south west, inner sep=0] (svg) at (0,0) {\includegraphics[width=0.24\textwidth]{grid.pdf}};
        \node[anchor=south west, inner sep=0] at (svg.south west) {\includegraphics[width=0.24\textwidth]{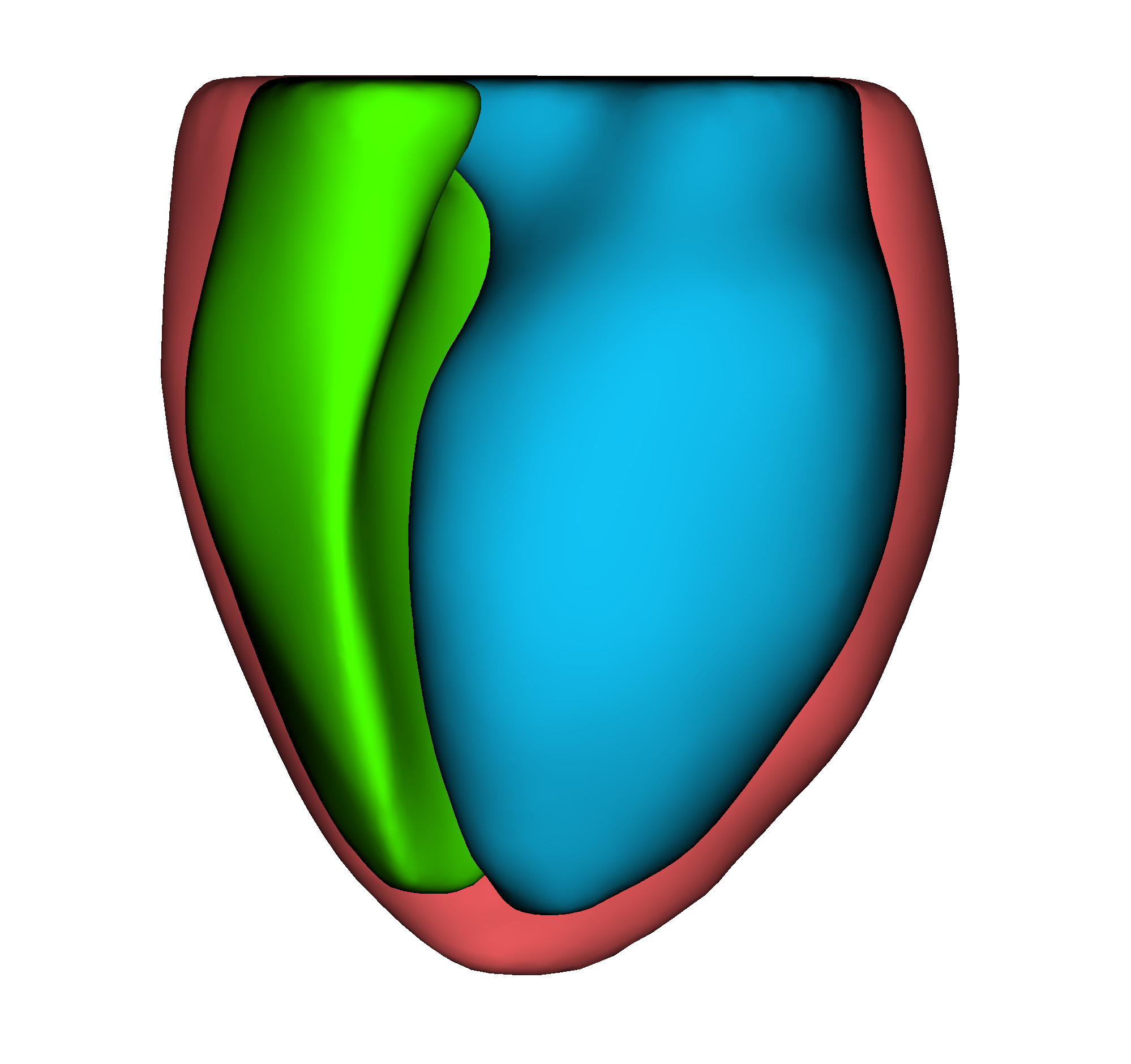}};
    \end{tikzpicture}
    \begin{tikzpicture}
        \node[anchor=south west, inner sep=0] (svg) at (0,0) {\includegraphics[width=0.24\textwidth]{grid.pdf}};
        \node[anchor=south west, inner sep=0] at (svg.south west) {\includegraphics[width=0.24\textwidth]{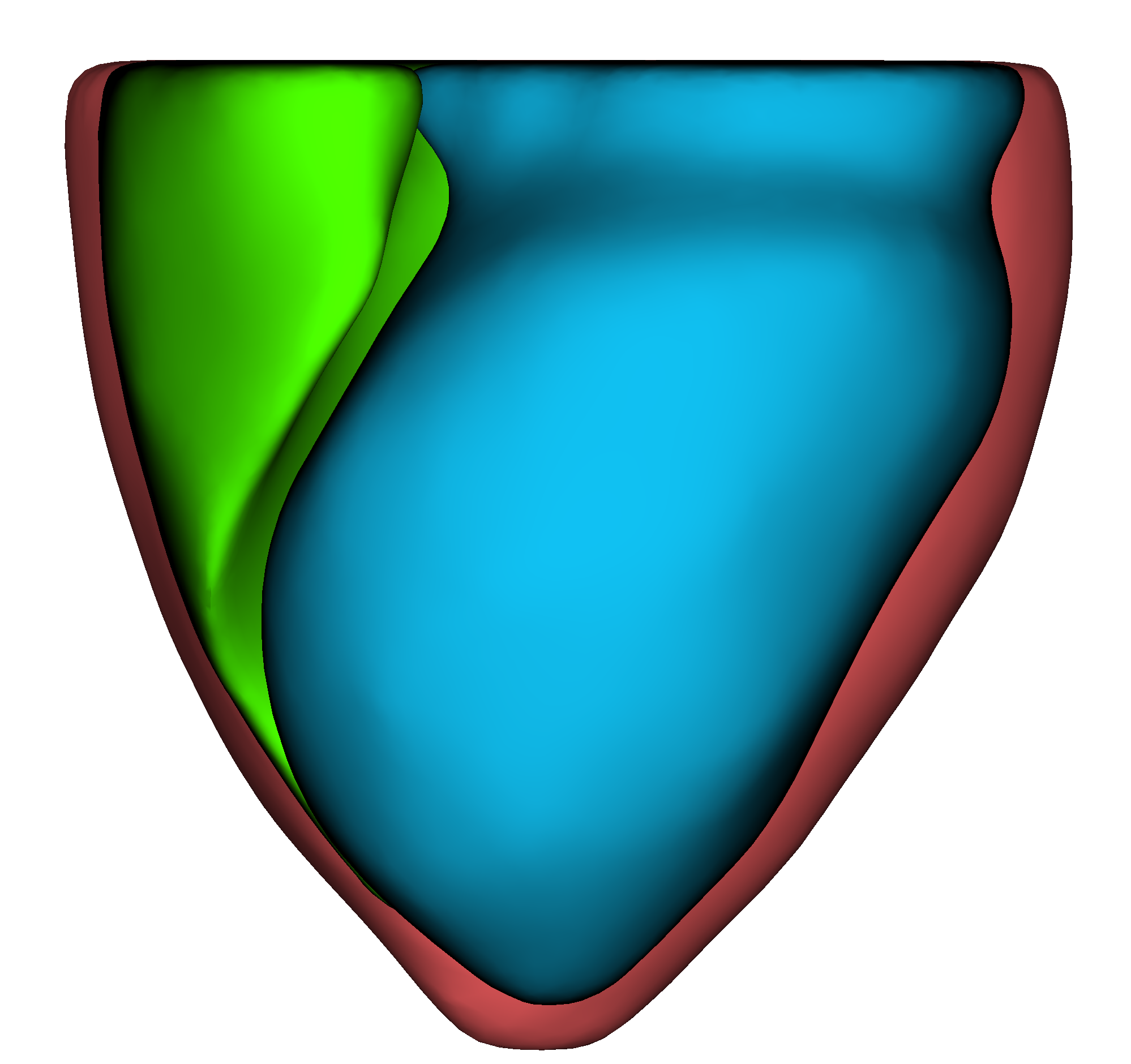}};
    \end{tikzpicture}

        \begin{tikzpicture}
        \node[anchor=south west, inner sep=0] (svg) at (0,0) {\includegraphics[width=0.24\textwidth]{grid.pdf}};
        \node[anchor=south west, inner sep=0] at (svg.south west) {\includegraphics[width=0.24\textwidth]{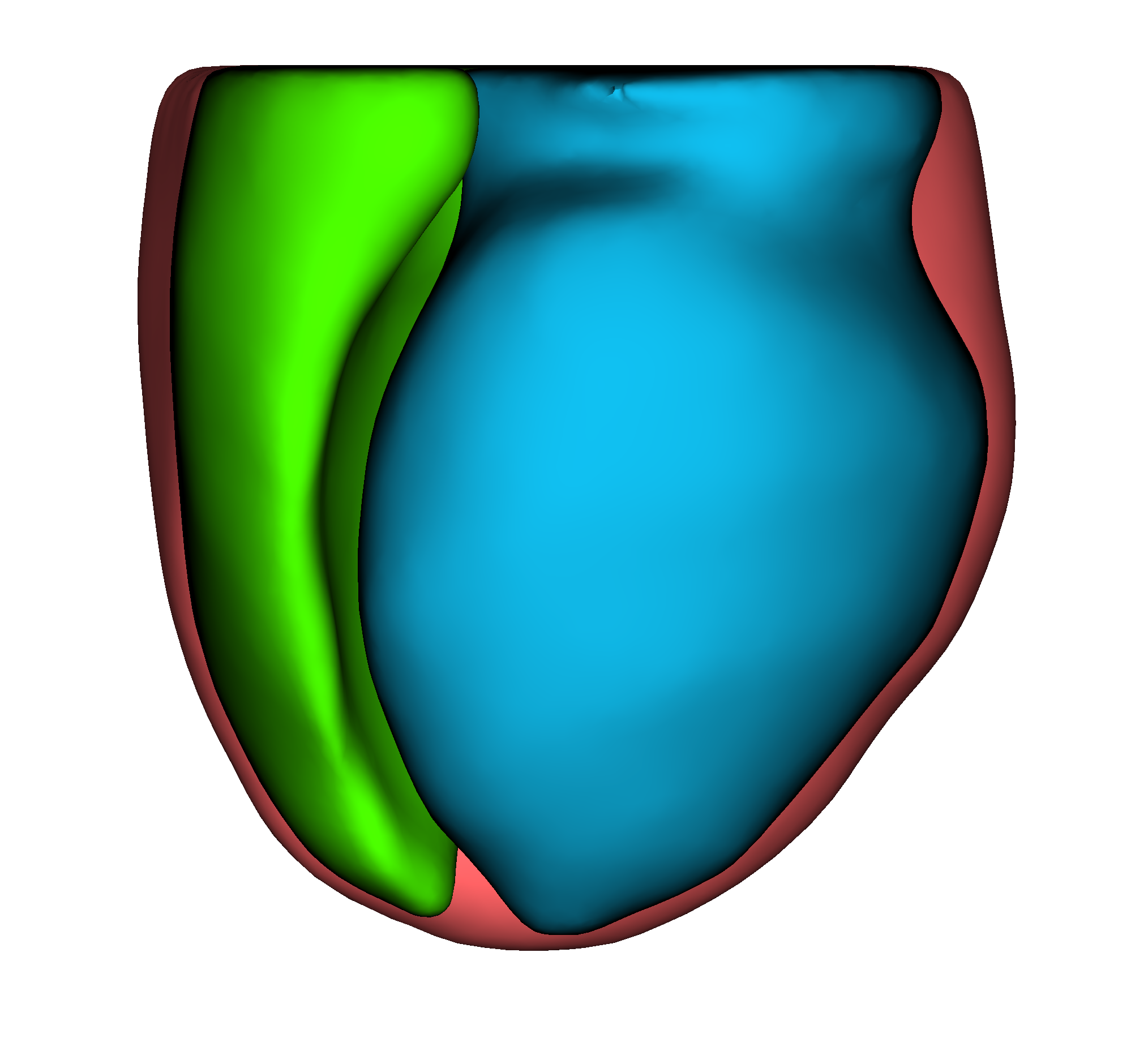}};
    \end{tikzpicture}
    \begin{tikzpicture}
        \node[anchor=south west, inner sep=0] (svg) at (0,0) {\includegraphics[width=0.24\textwidth]{grid.pdf}};
        \node[anchor=south west, inner sep=0] at (svg.south west) {\includegraphics[width=0.24\textwidth]{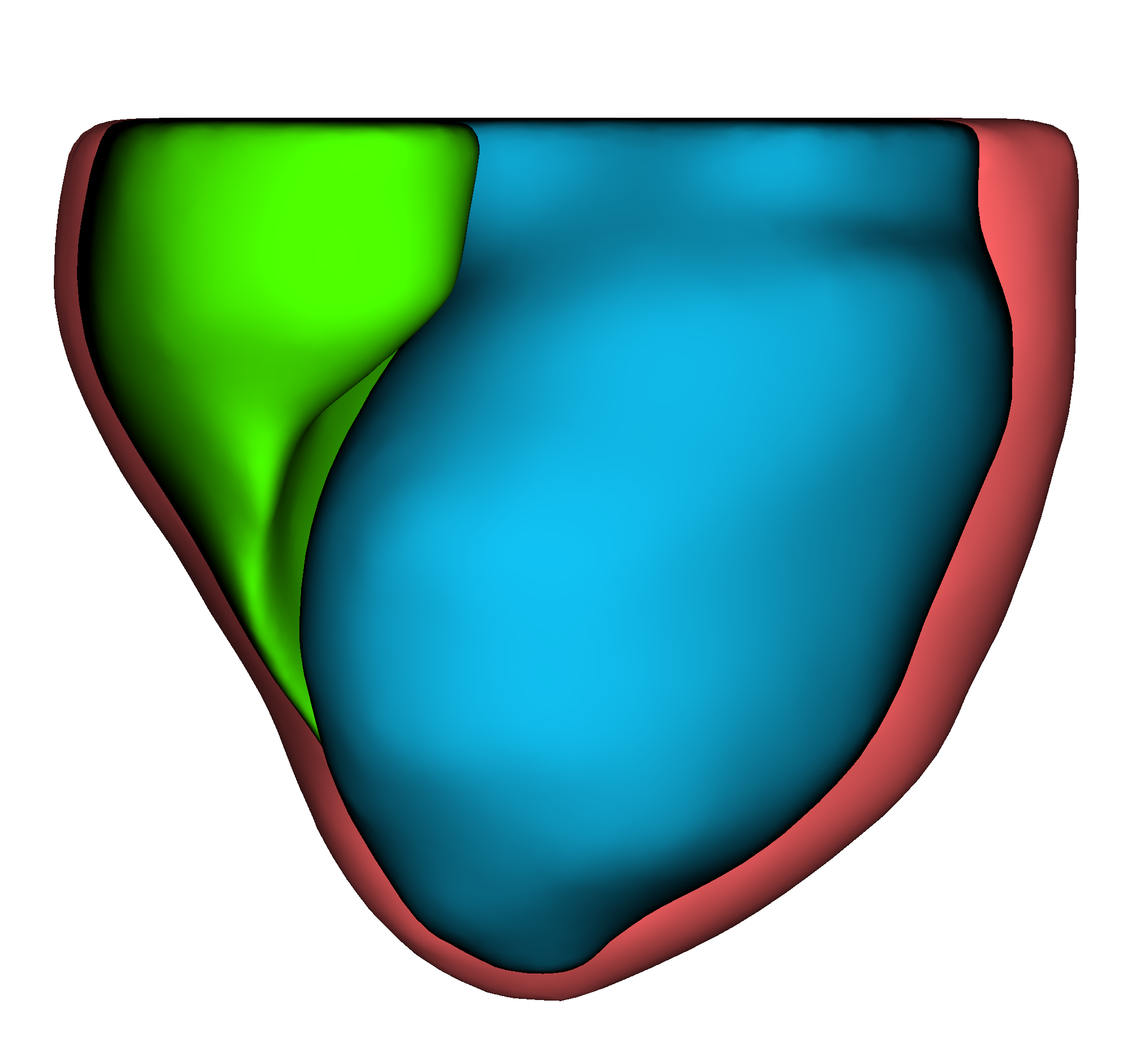}};
    \end{tikzpicture}
    \begin{tikzpicture}
        \node[anchor=south west, inner sep=0] (svg) at (0,0) {\includegraphics[width=0.24\textwidth]{grid.pdf}};
        \node[anchor=south west, inner sep=0] at (svg.south west) {\includegraphics[width=0.24\textwidth]{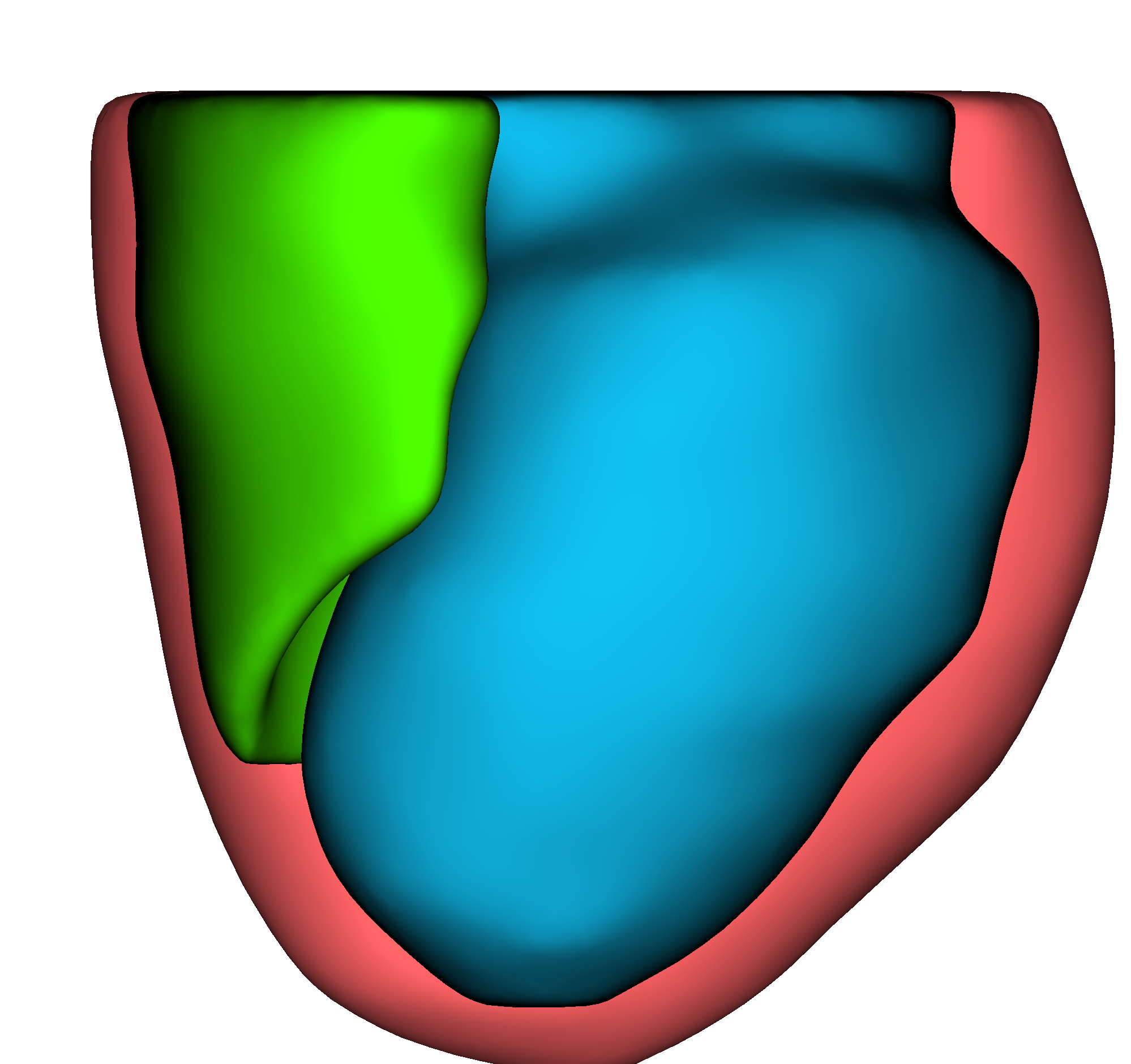}};
    \end{tikzpicture}
    \begin{tikzpicture}
        \node[anchor=south west, inner sep=0] (svg) at (0,0) {\includegraphics[width=0.24\textwidth]{grid.pdf}};
        \node[anchor=south west, inner sep=0] at (svg.south west) {\includegraphics[width=0.24\textwidth]{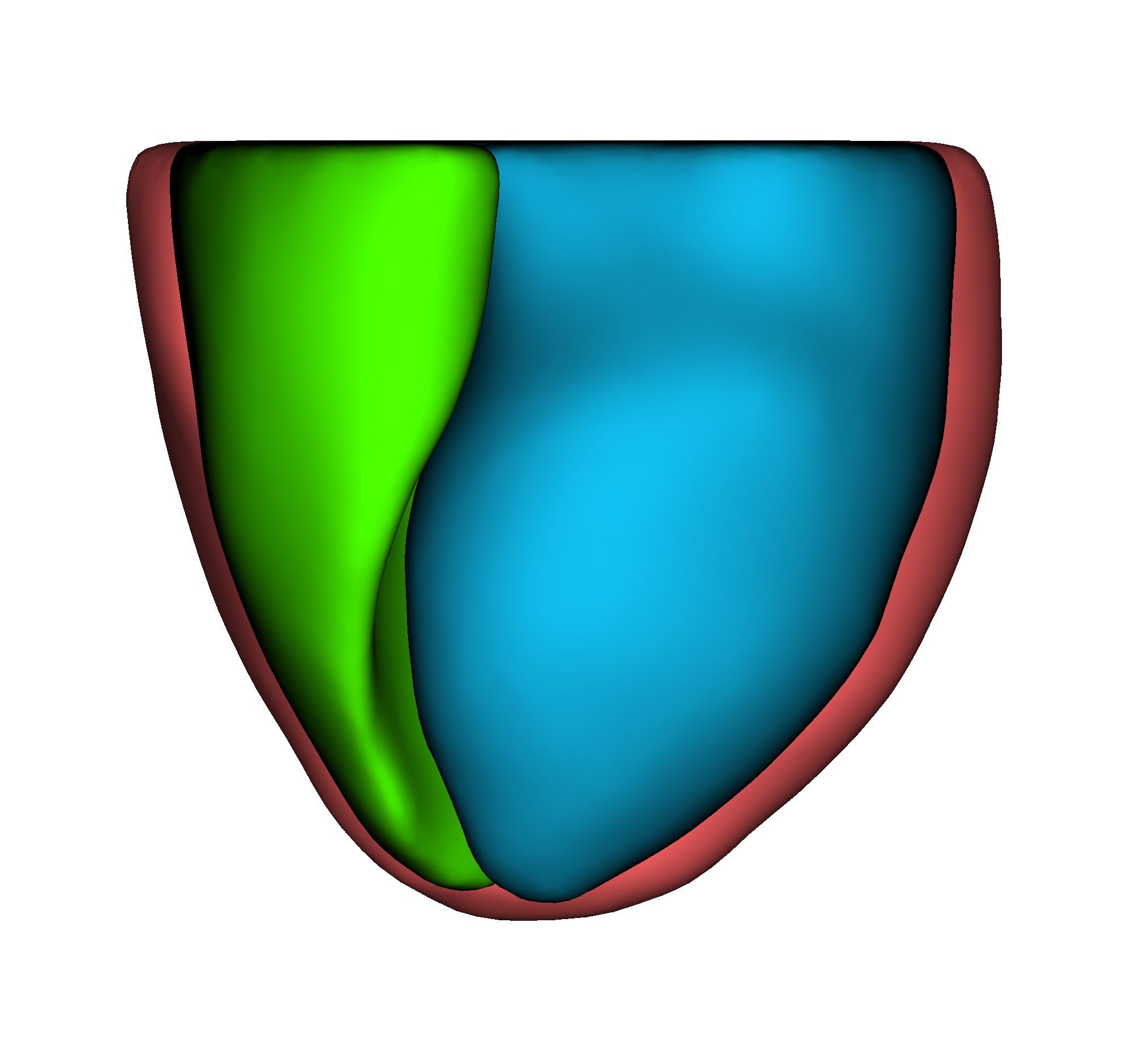}};
    \end{tikzpicture}
    \caption{\myhighlight{
       Exemplary heart shapes in frontal view from a cohort of 98 subjects. 
       Each heart comprises epicardial (red), left ventricular (blue), and right ventricular (green) luminal subsurfaces. 
       Substantial inter-subject variability is evident in global size and local features, such as right ventricular morphology.
       We use the top left geometry as the source.
    }}
    \label{fig:heart_set_examples}
\end{figure*}

\begin{figure}[h]
    \centering
    \includegraphics[width=0.8\textwidth]{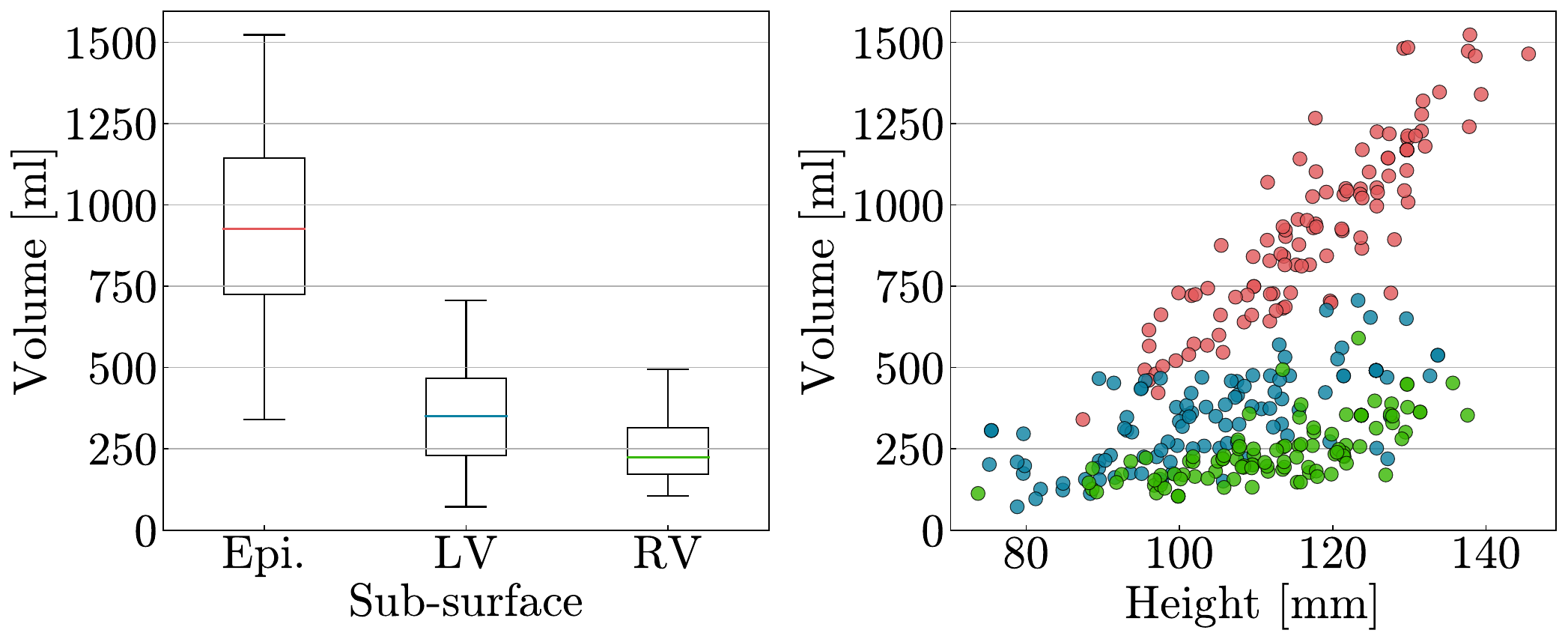}
    \caption{
    \myhighlight{
        Volume and height of the heart shape cohort. 
        On the left, each box represents the range from the first to the third quartile, with a colored line at the median. 
        The whiskers extend to the farthest data point within 150\% of the box range.
        On the right, the volume is scattered over the height.}
    }
    \label{fig:heart_cohort_metrics}
\end{figure}

\myhighlight{
We finally assess the two registration methods in a cohort setting by registering a single reference heart to a set of 98 patient-specific heart geometries.
Each heart shape again consists of epicardial, left, and right ventricular luminal subsurfaces, obtained from manual segmentation of CT images with an average voxel resolution of $1.2\text{ mm}$, resulting in a discretized surface with an average edge length of $6\text{ mm}$.
Fig.~\ref{fig:heart_set_examples} shows an excerpt of the pre-aligned heart cohort, and Fig.~\ref{fig:heart_cohort_metrics} displays the cohort's volume and height distribution.
Volume refers to the enclosed space of each subsurface, while the height corresponds to the subsurface’s extent along the anatomically defined long axis of the heart (i.e., the base-to-apex direction).
}

\FloatBarrier

\myhighlight{
To reduce inter-patient pose variability, all hearts are first rigidly aligned using three manually placed landmarks, followed by an Iterative Closest Point (ICP) refinement.
As source geometry, we select the heart whose total volume is closest to the mean volume of the cohort (top left in Fig.~\ref{fig:heart_set_examples}) and the remaining numerical setup for both methods is identical to the single-heart example in Sec.~\ref{sec:single_hearts}.
}

\begin{figure*}[ht]
    \centering
    \includegraphics[width=0.9\textwidth]{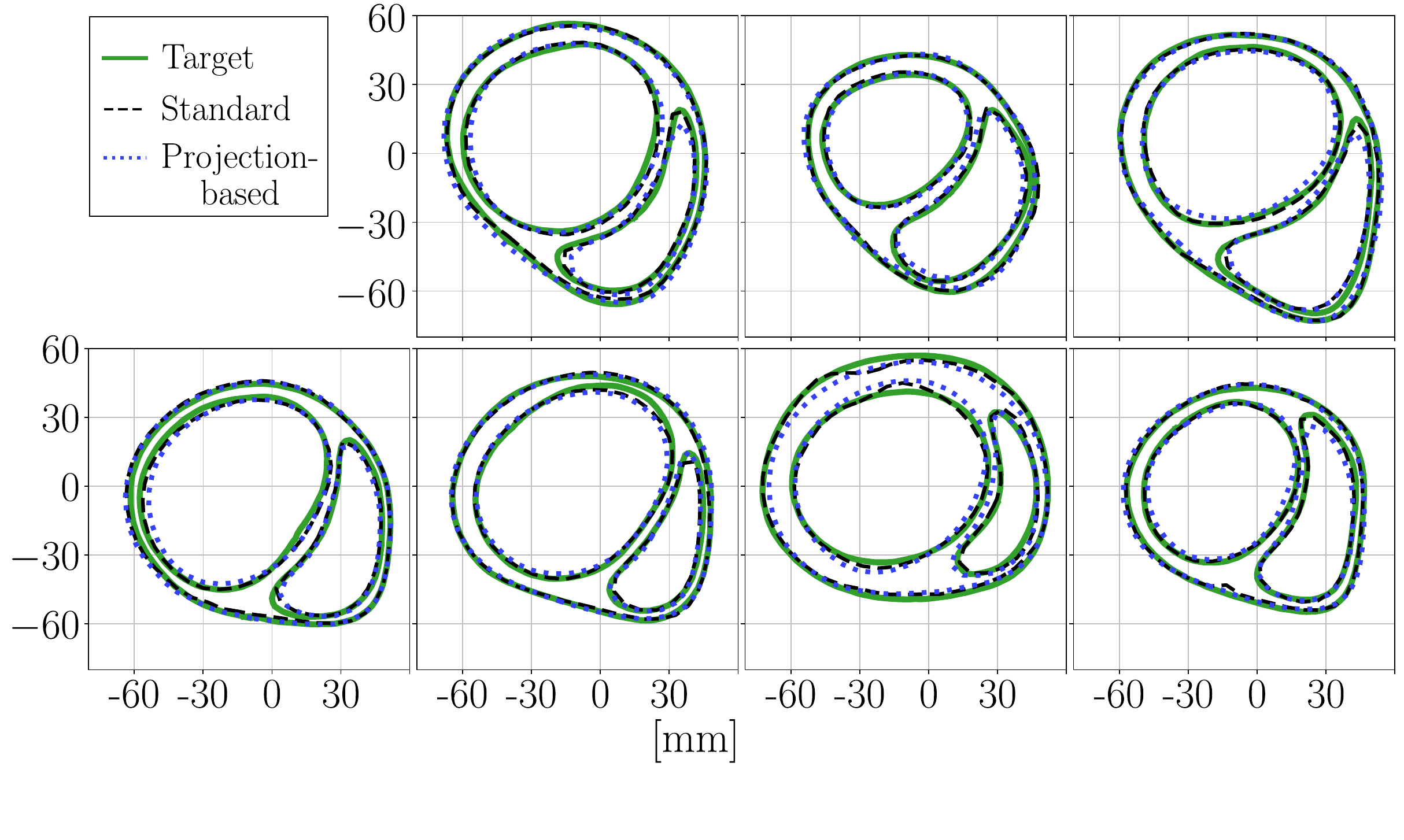}
    \caption{
    \myhighlight{
    Comparison of registration results between the standard and the projection-based method. 
    The hearts depicted in Fig.~\ref{fig:heart_set_examples} are sliced at 50\% of their respective height.
    Green depicts the target, and the dashed black and dotted blue lines depict the deformed source surfaces at $t=1$ of the standard and projection-based method, respectively.            
    }}
    \label{fig:cohort_slices}
\end{figure*}

\myhighlight{
Fig.~\ref{fig:cohort_slices} shows exemplary registration results for the geometries from Fig.~\ref{fig:heart_set_examples}, displaying the respective slice contours at 50\% of the corresponding target height.
The registration quality is quantified by the node-to-node distance.
Table~\ref{tab:results_heart_set} lists the cohort-wise mean and maximum node-to-node distances for each sub-surface.
Consistent with the previous example, the projection-based method yields slightly larger errors compared to the standard method, but the differences remain minor relative to the provided discretization and image resolution.
}

\begin{table}[h]
\myhighlightcolor
    \caption{
        \myhighlight{Heart cohort registration results.}
    }
    \begin{tabular}{lcc}
        \hline
        subsurface & Standard & Projection-based \\
         & method & method \\
        \hline
        \multicolumn{3}{c}{Mean nodal distance [mm]} \\
        \hline
        Epicardium &0.92& 1.34 \\
        Left ventricle &1.79&2.14\\
        Right ventricle &1.55&2.17\\
        Average &1.42& 1.88 \\
        \hline
        \multicolumn{3}{c}{Max. nodal distance [mm]} \\
        \hline
        Epicardium & 4.59 & 5.76  \\
        Left ventricle & 8.61 & 8.71\\
        Right ventricle & 8.42 & 10.10\\
        Average & 7.21 & 8.19 \\

        \hline
    \end{tabular}
    \label{tab:results_heart_set}
\end{table}

\myhighlight{
To analyze the residual RBM motion that is left even after the rigid pre-align, we extract the translational and rotational components in Table~\ref{tab:rbm_residual}.
}

\begin{table}[ht]
\myhighlightcolor
  \caption{\myhighlight{Heart cohort residual RBMs.}}
  \centering
  \begin{tabular}{l@{\hspace{16pt}}ccc@{\hspace{18pt}}cccc}
      \hline
      Component & Unit &x&y&z\\
      \hline
      \hline
      Translation & [mm] &  2.7 &  4.0 &  3.5   \\
      Rotation & [°] & 2.3 &  3.3 &  13.1 \\
      \hline
  \end{tabular}
  \label{tab:rbm_residual}
\end{table}
\FloatBarrier

\FloatBarrier

%% file: 4_discussion.tex
\section{Discussion and conclusion}

This work presents an extension to LDDMM surface matching that separates rigid body motion (RBMs) and shape-changing motion.
RBMs often arise in image and shape registration due to variations in imaging conditions and can hinder the establishment of physically meaningful spatial correspondences when intermixed with shape-changing motion.
By isolating shape-changing motion from the rigid transformations, the proposed method enables a focused analysis of each component individually.
Overall, the proposed method achieved similarly high registration quality.
We showed, for one example, that the additive decomposition of motion components preserves the surface mesh quality better.

To uniquely decompose motion into RBM and shape-changing components, a projector is introduced.
The projector is constructed from a basis representing translations and rotations.
In this study, all translations and rotations were removed.
However, the projector can be adapted to exclude only selected translation directions or rotation axes by modifying the projector basis \eqref{eq:projector_basis}.
Notably, when only translational components are considered, the projector remains constant over pseudo-time, eliminating the need for recomputation at each pseudo-time step.
The rotation basis vectors are derived from linearized rotations, which necessitates careful adjustment of the pseudo-time step to ensure sufficiently small rotation increments.

For simplicity and to provide an easily interpretable form, the RBM component was parameterized using pseudo-time-constant translation and a rotational component defined in terms of Euler's angles.
We note that Euler's angles can suffer from non-uniqueness, but, in our examples, this did not impede the solution of the surface matching problem.
Also, in our applications, we want to remove the RBM component to exclusively focus on the analysis of the shape-changing motion, which, of course, might be different in other settings.
\myhighlight{
    Alternatively, a rotation representation based on the Lie algebra so(3) or quaternions can be chosen at the cost of increased implementation complexity and reduced interpretability.
}

To assess the similarity between the registered surfaces, a surface matching metric identical to the standard LDDMM approach was employed in the objective function.
\myhighlight{
    However, the kinetic energy term was modified to be able to penalize RBM and shape-changing motion individually. 
    We chose to hardly penalize the RBM component to focus on the shape-changing motion, which is responsible for surface mesh distortion. 
    In the single pair of heart geometries example in Sec \ref{sec:single_hearts}, we showed that the proposed method can effectively reduce surface distortion while promoting a more uniform transformation.
}

Surface matching is a technique frequently used in various applications, particularly in conjunction with biomechanical models.
Accurately mapping organ geometries between different states is crucial for calibrating physical parameters.
To ensure adequate coverage of the region of interest, control points were initially distributed on a coarse grid encompassing the source surface.
More sophisticated control point placement and hyperparameter optimization strategies exist, but have not been discussed here.

\myhighlight{
    Our method resulted in a slightly higher node-to-node distance (Table \ref{tab:solution_table}) compared to the standard method, which may be attributed to neglecting the cross-terms in the kernel split (Section \ref{sec:first_mod}).
}
As the difference is significantly below the imaging resolution and segmentation accuracy, we consider it negligible for our purposes.

Due to the very high registration quality, the decomposed motion's flexibility, and the improved mesh preservation, the proposed approach is expected to assist shape analysis, especially for biomechanical problems.

%% file: 5_appendix.tex
\section{Subspace RKHS} \label{app:scalarproduct}

The Moore-Aronszajn theorem \cite{aronszajn_theory_1950} states that every symmetric, positive definite kernel defines a unique RKHS.

Here, we show these requirements for $\port$, but the same is true for $\ppar$.

Symmetry of $\Kort$ follows from the symmetry of $(\port)^T=\port$ and $\vK^T=\vK$:
\begin{align}
    {\Kort}^T 
    &= (\port\vK\port)^T 
    = {\port}^T\vK^T{\port}^T \\
    &= \port\vK\port = \vKort.
\end{align}
To show the positive definiteness, we assume that $\vK$ is positive definite and thus
\begin{equation}
    \boldsymbol{x}^{T} \vK \boldsymbol{x} \geq 0 \quad \forall \boldsymbol{x} \in \mathbf{V}.
\end{equation}
The equality holds if and only if \( \boldsymbol{x}=\boldsymbol{0} \).
Using the positive definiteness of $\vK$ the positive definiteness of $\vKort$ follows as
\begin{equation}
    \begin{aligned}
        \boldsymbol{x}^{T} \vKort \boldsymbol{x}
        &= \boldsymbol{x}^{T} \port \vK \port \boldsymbol{x} \\
        &= \boldsymbol{y}^{T} \vK \boldsymbol{y}
        \geq 0 \quad \forall \boldsymbol{y} \in \mathbf{V}^{\perp}.
    \end{aligned}
\end{equation}
The above equality holds true exclusively if $\boldsymbol{y}= \port \boldsymbol{x}=\boldsymbol{0}$. This is the case if $\boldsymbol{x} \in V^\parallel$, therefore $\boldsymbol{x} \notin V^\perp$, or $\boldsymbol{x} = \boldsymbol{0}$. $V^\perp$ can be completed into a RKHS characterized by $\vKort$ by considering possible infinite sums of terms in \eqref{eq:arbitrary_motion}.

Additionally we show that the induced scalar product is valid by showing its symmetry, positive definiteness, and linearity in both arguments.
The scalar product of two vector fields $\vv = \vK ( \cdot , \vcp )\vmom$ and $\vv' = \vK ( \cdot , \vcp' )\boldsymbol{\vmom'}$ from \eqref{eq:scalaprod} is defined as
\begin{equation}
    \langle \vv, \vv' \rangle_{V^{\perp}}
    = \vmom^{T} \vKort(\vcp, \vcp') \boldsymbol{\vmom'}.
\end{equation}
Symmetry of the scalar product follows from the symmetry of $\vKort$
\begin{equation}
    \begin{aligned}
        \langle \vv, \vv' \rangle_{V^{\perp}}
        &= \vmom^{T} \vKort(\vcp, \vcp') \boldsymbol{\vmom'} \\
        &= \left[\boldsymbol{\vmom'}^T \left(\vKort(\vcp, \vcp') \right)^T \vmom\right]^T\\
        &= \langle \vv', \vv \rangle_{V^{\perp}}.
    \end{aligned}
\end{equation}
To show linearity in the first argument, we use another arbitrary element of the vector space $\vv_{3}=\vK ( \cdot , \vcp )\boldsymbol{\gamma} \in V$ and a scalar $\lambda \in \R$, therefore,
\begin{equation}
    \begin{aligned}
        &\langle \vv + \lambda\vv', \vv_{3} \rangle_{V^{\perp}}\\
        &= \left(\vmom + \lambda\boldsymbol{\vmom'} \right)^{T} \vKort(\vcp, \vcp') \boldsymbol{\gamma} \\
        &=  \vmom^{T} \vKort(\vcp, \vcp') \boldsymbol{\gamma}
        +  \lambda\boldsymbol{\vmom'}^{T} \vKort(\vcp, \vcp') \boldsymbol{\gamma} \\
        &= \langle \vv, \vv_{3} \rangle_{V^{\perp}}
        + \lambda\langle \vv', \vv_{3} \rangle_{V^{\perp}} \quad \forall \lambda \in \R.
    \end{aligned}
\end{equation}
Linearity with respect to the second argument can be shown analogously.

\section{Momenta evolution} \label{app:mom_evolution}
\myhighlight{
The motion $\dot{\vcp}$ must minimize the kinetic energy $\tilde{E} = \gamma^\perp E^\perp + \gamma^\parallel E^\parallel$ to ensure uniqueness of motion. Therefore, the condition 
\begin{equation} \label{eq:e_kin_derivative}
     \frac{d \tilde{E}}{d\vmom}
     = \gamma^\perp \frac{d E^\perp}{d\vmom} + \gamma^\parallel \frac{d E^\parallel}{d\vmom} 
     =\boldsymbol{0},
\end{equation}
must be fulfilled. The derivation of the summands follow the same steps as in Appendix A of \cite{durrleman_morphometry_2014}, and result in
\begin{gather}
        \frac{d E^\perp}{d\vmom} 
        =\int_{0}^{1}\left(-2 \dvmom^{T}- \vmom^{T}\frac{\partial \dot{\vcp}^{\perp}}{\partial \vcp} \right)\frac{d \vcp}{d \vmom} \:dt,\\
        \frac{d E^\parallel}{d\vmom} 
        =\int_{0}^{1}\left(-2 \dvmom^{T}- \vmom^{T}\frac{\partial \dot{\vcp}^{\parallel}}{\partial \vcp} \right)\frac{d \vcp}{d \vmom} \:dt.
\end{gather}
Therefore, \eqref{eq:e_kin_derivative} is fulfilled if
\begin{equation}
    \begin{aligned}
        \dot{\vmom} 
        &= \dfrac{-1}{2\gamma} 
        \left[ 
        \gamma^\perp \left(\frac{\partial \dot{\vcp}^{\perp}}{\partial \vcp}\right)^T 
        + \gamma^\parallel \left(\frac{\partial \dot{\vcp}^{\parallel}}{\partial \vcp}\right)^T
        \right] \vmom, 
    \end{aligned}
\end{equation}
where we used $\gamma = \gamma^\perp + \gamma^\parallel$.
}

\section{Gradient of the rigid body mode free Kernel} \label{app:kernel_gradient}

The gradient  $\nabla_{\vcp} \vKort$ with respect to the current control point positions $\vcp$ is needed in equation \eqref{eq:geo_mom_new}.
For brevity, we use $\nabla := \nabla_{\vcp}$ in

\begin{equation}
    \begin{aligned}
        \nabla \vKort
        &= \nabla \left( \port \vK \port \right) \\
        &= \nabla\port \vK \port \\
        &\quad + \port \nabla\vK \port \\
        &\quad + \port \vK \nabla\port.
    \end{aligned}
\end{equation}
The gradient of the kernel
\begin{equation}
    \begin{aligned}
        \nabla\vK
        &= \frac{d}{d\cp_j} k(\cp_i, \cp_k) \\
        &= \frac{2}{\sigma^{2}_V} (\delta_{jk}-\delta_{ik}) k(\cp_i, \cp_k) \left| \cp_i - \cp_k \right|
    \end{aligned}
\end{equation}
is known, see \cite{durrleman_morphometry_2014}, where $\delta_{jk}$ is the Kronecker delta and Einstein summation convention applies.
The gradient of the projector is
\begin{equation}
    \begin{aligned}
        \nabla\port
        &= \nabla \left( \eye -  \basis \vA^{-1}\basis^T\right) \\
        &= \nabla \basis \vA^{-1}\basis^T \\
        &\quad + \basis \nabla\vA^{-1}\basis^T \\
        &\quad + \basis\vA^{-1}\nabla\basis^T.
    \end{aligned}
\end{equation}
We first concentrate on the term $\nabla\vA^{-1}$ in the second summand.
We can write $\eye=\vA\vA^{-1}$ and know $\nabla\eye=\boldsymbol{0}$.
Rearranging
\begin{equation}
    \begin{aligned}
        \nabla\eye
        &=\nabla\left(\vA\vA^{-1}\right) \\
        &=\nabla \vA \vA^{-1} \\
        &\quad + \vA \nabla \vA^{-1} \\
        &= \boldsymbol{0}
    \end{aligned}
\end{equation}
gives
\begin{equation}
    \nabla \vA^{-1} = - \vA^{-1} \nabla \vA \vA^{-1},
\end{equation}
with
\begin{equation}
    \nabla \vA = \nabla\basis^T \basis + \basis^T \nabla\basis.
\end{equation}
The only remaining unknown term is the gradient of the basis $\basis$ itself.
As $\basis$ was defined in \ref{eq:projector_basis_stack} as $\basis = \left[\hat{\basis}_1,\hat{\basis}_2,\hdots,\hat{\basis}_{N_x}\right]^T$, its gradient is
\begin{equation}
    \begin{aligned}
        \nabla \basis
        &= \frac{d}{dc_{jJ}}\left(\hat{\basis}_{i}\right) \\
        &= \delta_{ij}
        \left[\begin{array}{cccccc}
            0 & 0 & 0 & 0            & \delta_{3J}  & -\delta_{2J} \\
            0 & 0 & 0 & -\delta_{3J} & 0            & \delta_{1J}  \\
            0 & 0 & 0 & \delta_{2J}  & -\delta_{1J} & 0
        \end{array}\right],
    \end{aligned}
\end{equation}
where $i,j = 1,\dots,N_{cp}$ and $J=1,2,3$.

\section{Subspace projection}  \label{app:subProj}
For completeness and clarity, we give a short repetition of subspace projections.
See, e.g., \cite{banerjee_linear_2014} for elaborate details.
The motion is contained in the finite-dimensional space $V$.
Let $V^\parallel$ and $V^\perp$ be subspaces of $V$, such that $V = V^{\perp} \oplus V^{\parallel}$ and $V^\parallel \cap V^\perp = \{0\}$.
Then, for each element $\boldsymbol{v} \in V$, there exists a unique decomposition
\begin{equation} \label{eq:vector_split}
    \vv= \vv^\parallel + \vv^\perp,
\end{equation}
where $ \vv^\parallel \in V^\parallel$ and  $\vv^\perp \in V^\perp$.
The elements of the subspaces can be computed by introducing a projection $P^\parallel:V \rightarrow V^\parallel$ and its orthogonal complement  $P^\perp:V \rightarrow V^\perp$, defined as $P^\perp=I-P^\parallel$.
Both projections are linear transformations such that  $\left(P^\parallel\right)^2 = P^\parallel$ and $\left(P^\perp\right)^2 = P^\perp$.
The projection $P^\parallel$ is called orthogonal if its nullspace is orthogonal to its image.
Their dimension is $\proj^\parallel, \proj^\perp \in \R^{3N_x \times 3N_x}$.

Based on these projections, we define two orthogonal projectors as idempotent and symmetric matrices $\proj^\parallel$ and $\proj^\perp$, with the following properties
\begin{equation}
    \proj^\parallel \proj^\perp = \boldsymbol{0}, \quad \proj^\parallel + \proj^\perp = \eye.
\end{equation}
Then, the subspace elements from \eqref{eq:vector_split} are defined as
\begin{align}
    \vv^\parallel &= \proj^\parallel \vv \\
    \vv^\perp &= \proj^\perp \vv.
\end{align}

\section{Elemental rotation matrices}  \label{app:rotmats}

The matrices of the three elemental rotations are defined as
\begin{equation}
    \begin{aligned}
    \hat{\boldsymbol{R}}_{x}(\rotangle_{x}) =
    \left[\begin{array}{ccc}
    1 & 0 & 0 \\
    0 & \cos(\rotangle_{x}) & -\sin(\rotangle_{x}) \\
    0 & \sin(\rotangle_{x}) & \cos(\rotangle_{x})
    \end{array}\right]& \\
    \hat{\boldsymbol{R}}_{y}(\rotangle_{y}) =
    \left[\begin{array}{ccc}
        \cos(\rotangle_{y}) & 0 & -\sin(\rotangle_{y}) \\
        0 & 1 & 0 \\
        \sin(\rotangle_{y}) & 0 & \cos(\rotangle_{y})
    \end{array}\right]& \\
    \hat{\boldsymbol{R}}_{z}(\rotangle_{z}) =
    \left[\begin{array}{ccc}
        \cos(\rotangle_{z}) & -\sin(\rotangle_{z}) & 0 \\
        \sin(\rotangle_{z}) & \cos(\rotangle_{z}) & 0 \\
        0 & 0 & 1
    \end{array}\right]&.
    \end{aligned}
\end{equation}

\clearpage
\myhighlight{
\section{Pseudo-code} \label{app:pseudocode}
\begin{figure*}[ht]
\centering
\begin{minipage}{\textwidth}
\begin{algorithm}[H]
\myhighlightcolor
\caption{Surface matching with decomposed motion}
\label{alg:pseudo_code}
\begin{algorithmic}
        \State \textbf{Input:} source $\mathcal{S}$, target $\mathcal{T}$, control points $\vcp_0$, hyperparameters $\sigma_V, \sigma_W, \gamma^\perp, \gamma^\parallel$
        \State \textbf{Output:} initial momenta $\vmom_0$, translations $\hat{\vtrans}_0$, rotation angles $\vangle_0$
        \State (Optionally) align source $\mathcal{S}$ and target $\mathcal{T}$ rigidly
        \State Initialize $[\vmom_{0}, \hat{\vtrans}_{0}, \vangle_{0}]^T = \boldsymbol{0}$
        \While{ not converged }
        \State Reset $t = 0$
        \State Reset $\boldsymbol{x} = \boldsymbol{x}_0, \vcp = \vcp_0, \vmom = \vmom_0$
        \State Reset $\vrot_{\boldsymbol{x}}=\vrot_{\boldsymbol{x}}(\boldsymbol{x}_0, \vangle_0), \vrot_{\vcp} = \vrot_{\vcp}(\vcp_0, \vangle_0)$
        \State Compute translation fields $\vtrans_{\vcp}(\hat{\vtrans}_{0}), \vtrans_{\boldsymbol{x}}(\hat{\vtrans}_{0})$
        \Comment The translation fields are constant in pseudo-time.
        \While{$t < 1$}
            \State Compute Projectors $\port(\vcp), \port(\boldsymbol{x})$
            \Comment See \eqref{eq:projector_definition}
            \State Compute Kernels $\vKort(\boldsymbol{x}, \vcp), \vKort(\vcp, \vcp), \nabla_{\vcp}  \vKort(\vcp, \vcp)$
            \Comment See \eqref{eq:kernel_split} and App.~\ref{app:kernel_gradient}
            \State Compute rotation fields $\vrot_{\boldsymbol{x}}, \vrot_{\vcp}$ 
            \Comment See \eqref{eq:rotation_field}
            \State Update $\vcp, \vmom, \boldsymbol{x}, t$ 
            \Comment With Runge-Kutta scheme or similar
        \EndWhile
        \State Compute similarity metric $A(\mathcal{S}, \mathcal{T})\Big|_{t=1}$
        \Comment See \eqref{eq:surface_metric} or \cite{durrleman_morphometry_2014,charon_varifold_2013}
        \State Compute kinetic Energy $\tilde{E} = \gamma^\perp E^\perp + \gamma^\parallel E^\parallel$
        \Comment See \eqref{eq:new_e_kin} and \eqref{eq:kin_e_parallel_new}
        \State Update $\vmom_{0}, \hat{\vtrans}_{0}, \vangle_{0}$
        \Comment With L-BFGS or similar
        \EndWhile
\end{algorithmic}
\end{algorithm}
\end{minipage}
\end{figure*}
}

\FloatBarrier

%% file: main.bbl
\begin{thebibliography}{32}
\providecommand{\natexlab}[1]{#1}
\providecommand{\url}[1]{{#1}}
\providecommand{\urlprefix}{URL }
\providecommand{\doi}[1]{\url{https://doi.org/#1}}
\providecommand{\eprint}[2][]{\url{#2}}
 \bibcommenthead

\bibitem[{Antonsanti et~al.(2021)Antonsanti, Glaun{\`e}s, Benseghir, Jugnon, and Kaltenmark}]{antonsanti_partial_2021}
Antonsanti PL, Glaun{\`e}s J, Benseghir T, et~al (2021) Partial {{Matching}} in the {{Space}} of {{Varifolds}}. In: Feragen A, Sommer S, Schnabel J, et~al (eds) Information {{Processing}} in {{Medical Imaging}}, vol 12729. Springer International Publishing, Cham, p 123--135, \doi{10.1007/978-3-030-78191-0_10}

\bibitem[{Arguill{\`e}re et~al.(2016{\natexlab{a}})Arguill{\`e}re, Miller, and Younes}]{arguillere_diffeomorphic_2016}
Arguill{\`e}re S, Miller MI, Younes L (2016{\natexlab{a}}) Diffeomorphic {{Surface Registration}} with {{Atrophy Constraints}}. SIAM journal on imaging sciences 9(3):975--1003. \doi{10.1137/15m104431x}

\bibitem[{Arguill{\`e}re et~al.(2016{\natexlab{b}})Arguill{\`e}re, Tr{\'e}lat, Trouv{\'e}, and Younes}]{arguillere_registration_2016}
Arguill{\`e}re S, Tr{\'e}lat E, Trouv{\'e} A, et~al (2016{\natexlab{b}}) Registration of {{Multiple Shapes}} using {{Constrained Optimal Control}}. SIAM Journal on Imaging Sciences 9(1):344--385. \doi{10.1137/15M1006726}

\bibitem[{Aronszajn(1950)}]{aronszajn_theory_1950}
Aronszajn N (1950) Theory of reproducing kernels. Transactions of the American Mathematical Society 68(3):337--404. \doi{10.1090/S0002-9947-1950-0051437-7}

\bibitem[{Banerjee and Roy(2014)}]{banerjee_linear_2014}
Banerjee S, Roy A (2014) Linear Algebra and Matrix Analysis for Statistics. Texts in Statistical Science, CRC Press, Taylor \& Francis, Boca Raton Fla.

\bibitem[{Beg et~al.(2005)Beg, Miller, Trouv{\'e}, and Younes}]{beg_computing_2005-1}
Beg MF, Miller MI, Trouv{\'e} A, et~al (2005) Computing {{Large Deformation Metric Mappings}} via {{Geodesic Flows}} of {{Diffeomorphisms}}. International Journal of Computer Vision 61(2):139--157. \doi{10.1023/B:VISI.0000043755.93987.aa}

\bibitem[{Ben~Amor et~al.(2022)Ben~Amor, Arguillere, and Shao}]{ben_amor_resnet-lddmm_2022}
Ben~Amor B, Arguillere S, Shao L (2022) {{ResNet-LDDMM}}: {{Advancing}} the {{LDDMM Framework}} using {{Deep Residual Networks}}. IEEE Transactions on Pattern Analysis and Machine Intelligence pp 1--1. \doi{10.1109/TPAMI.2022.3174908}

\bibitem[{B{\^o}ne et~al.(2018)B{\^o}ne, Louis, Martin, and Durrleman}]{reuter_deformetrica_2018}
B{\^o}ne A, Louis M, Martin B, et~al (2018) Deformetrica 4: {{An Open-Source Software}} for {{Statistical Shape Analysis}}. In: Reuter M, Wachinger C, Lombaert H, et~al (eds) Shape in {{Medical Imaging}}, vol 11167. Springer International Publishing, Cham, pp 3--13, \doi{10.1007/978-3-030-04747-4_1}

\bibitem[{Braga et~al.(2019)Braga, Zimmer, Dumoncel, Samir, {de Beer}, Zanolli, Pinto, Rohlf, and Grine}]{braga_efficacy_2019}
Braga J, Zimmer V, Dumoncel J, et~al (2019) Efficacy of diffeomorphic surface matching and {{3D}} geometric morphometrics for taxonomic discrimination of {{Early Pleistocene}} hominin mandibular molars. Journal of Human Evolution 130:21--35. \doi{10.1016/j.jhevol.2019.01.009}

\bibitem[{Burden et~al.(2016)Burden, Burden, and Faires}]{burden_numerical_2016}
Burden A, Burden R, Faires J (2016) Numerical {{Analysis}}, 10th Ed. Cengage Learning, \doi{10.13140/2.1.4830.2406}

\bibitem[{Ceritoglu et~al.(2009)Ceritoglu, Oishi, Li, Chou, Younes, Albert, Lyketsos, {van Zijl}, Miller, and Mori}]{ceritoglu_multi-contrast_2009}
Ceritoglu C, Oishi K, Li X, et~al (2009) Multi-contrast large deformation diffeomorphic metric mapping for diffusion tensor imaging. NeuroImage 47(2):618--627. \doi{10.1016/j.neuroimage.2009.04.057}

\bibitem[{Charon and Trouv{\'e}(2013)}]{charon_varifold_2013}
Charon N, Trouv{\'e} A (2013) The varifold representation of non-oriented shapes for diffeomorphic registration. SIAM Journal on Imaging Sciences 6(4):2547--2580. \doi{10.1137/130918885}, {\href{https://arxiv.org/abs/1304.6108}{{arXiv:1304.6108}}} {[cs, math]}

\bibitem[{Durrleman et~al.(2009)Durrleman, Pennec, Trouv{\'e}, and Ayache}]{durrleman_statistical_2009}
Durrleman S, Pennec X, Trouv{\'e} A, et~al (2009) Statistical models of sets of curves and surfaces based on currents. Medical Image Analysis 13(5):793--808. \doi{10.1016/j.media.2009.07.007}

\bibitem[{Durrleman et~al.(2014)Durrleman, Prastawa, Charon, Korenberg, Joshi, Gerig, and Trouv{\'e}}]{durrleman_morphometry_2014}
Durrleman S, Prastawa M, Charon N, et~al (2014) Morphometry of anatomical shape complexes with dense deformations and sparse parameters. NeuroImage 101:35--49. \doi{10.1016/j.neuroimage.2014.06.043}

\bibitem[{Glaun{\`e}s et~al.(2008)Glaun{\`e}s, Qiu, Miller, and Younes}]{glaunes_large_2008}
Glaun{\`e}s J, Qiu A, Miller MI, et~al (2008) Large {{Deformation Diffeomorphic Metric Curve Mapping}}. International Journal of Computer Vision 80(3):317--336. \doi{10.1007/s11263-008-0141-9}

\bibitem[{Glaun{\`e}s(2005)}]{glaunes_thesis_2005}
Glaun{\`e}s JA (2005) Transport par diff\'eomorphismes de points, de mesures et de courants pour la comparaison de formes et l'anatomie num\'erique. PhD thesis, University of Paris

\bibitem[{Gris et~al.(2018)Gris, Durrleman, and Trouv{\'e}}]{gris_subriemannian_2018}
Gris B, Durrleman S, Trouv{\'e} A (2018) A {{Sub-Riemannian Modular Framework}} for {{Diffeomorphism-Based Analysis}} of {{Shape Ensembles}}. SIAM Journal on Imaging Sciences 11(1):802--833. \doi{10.1137/16M1076733}

\bibitem[{Guan et~al.(2024)Guan, Tian, Li, and Gao}]{guan_using_2024}
Guan D, Tian L, Li W, et~al (2024) Using {{LDDMM}} and a kinematic cardiac growth model to quantify growth and remodelling in rat hearts under {{PAH}}. Computers in Biology and Medicine 171:108218. \doi{10.1016/j.compbiomed.2024.108218}

\bibitem[{Guigui and Pennec(2022)}]{guigui_parallel_2022}
Guigui N, Pennec X (2022) Parallel transport, a central tool in geometric statistics for computational anatomy: {{Application}} to cardiac motion modeling. In: Nielsen F, Srinivasa~Rao ASR, Rao CR (eds) Handbook of {{Statistics}}, Geometry and {{Statistics}}, vol~46, 1st edn. Elsevier, p 285--326, \doi{10.1016/bs.host.2022.03.006}

\bibitem[{Hsieh et~al.(2021)Hsieh, Arguill{\`e}re, Charon, and Younes}]{hsieh_diffeomorphic_2021}
Hsieh DN, Arguill{\`e}re S, Charon N, et~al (2021) Diffeomorphic shape evolution coupled with a reaction-diffusion {{PDE}} on a growth potential. \doi{10.48550/ARXIV.2101.06508}

\bibitem[{Lin et~al.(2022)Lin, Zhao, and Xing}]{lin_asymmetry_2022}
Lin S, Zhao Y, Xing S (2022) Asymmetry of {{Endocast Surface Shape}} in {{Modern Humans Based}} on {{Diffeomorphic Surface Matching}}. Symmetry 14(7):1459. \doi{10.3390/sym14071459}

\bibitem[{Meyer(2000)}]{meyer_matrix_2000}
Meyer CD (2000) Matrix Analysis and Applied Linear Algebra. {Society for Industrial and Applied Mathematics}, Philadelphia

\bibitem[{Nocedal(1980)}]{nocedal_updating_1980}
Nocedal J (1980) Updating quasi-{{Newton}} matrices with limited storage. Mathematics of Computation 35(151):773--782. \doi{10.1090/S0025-5718-1980-0572855-7}

\bibitem[{Polzin(2018)}]{polzin_large_2018}
Polzin T (2018) Large {{Deformation Diffeomorphic Metric Mappings}}. PhD thesis, University of L\"ubeck, L\"ubeck

\bibitem[{Runge(1895)}]{runge_ueber_1895}
Runge C (1895) {Ueber die numerische Aufl\"osung von Differentialgleichungen}. Mathematische Annalen 46(2):167--178. \doi{10.1007/BF01446807}

\bibitem[{Sommer et~al.(2012)Sommer, Nielsen, Darkner, and Pennec}]{sommer_higherorder_2012}
Sommer S, Nielsen M, Darkner S, et~al (2012) Higher-{{Order Momentum Distributions}} and {{Locally Affine LDDMM Registration}}. \doi{10.48550/arXiv.1112.3166}, {\href{https://arxiv.org/abs/1112.3166}{{arXiv:1112.3166}}}

\bibitem[{Stouffer et~al.(2022)Stouffer, Witter, Tward, and Miller}]{stouffer_projective_2022}
Stouffer KM, Witter MP, Tward DJ, et~al (2022) Projective diffeomorphic mapping of molecular digital pathology with tissue {{MRI}}. Communications Engineering 1(1):1--11. \doi{10.1038/s44172-022-00044-1}

\bibitem[{Vaillant and Glaun{\`e}s(2005)}]{hutchison_surface_2005}
Vaillant M, Glaun{\`e}s J (2005) Surface {{Matching}} via {{Currents}}. In: Hutchison D, Kanade T, Kittler J, et~al (eds) Information {{Processing}} in {{Medical Imaging}}, vol 3565. Springer Berlin Heidelberg, Berlin, Heidelberg, p 381--392, \doi{10.1007/11505730_32}

\bibitem[{Vanvek et~al.(2001)Vanvek, Brezina, and Mandel}]{vanvek_convergence_2001}
Vanvek P, Brezina M, Mandel J (2001) Convergence of algebraic multigrid based on smoothed aggregation. Numerische Mathematik 88(3):559--579. \doi{10.1007/s211-001-8015-y}

\bibitem[{Vialard et~al.(2012)Vialard, Risser, Rueckert, and Cotter}]{vialard_diffeomorphic_2012-1}
Vialard FX, Risser L, Rueckert D, et~al (2012) Diffeomorphic {{3D Image Registration}} via {{Geodesic Shooting Using}} an {{Efficient Adjoint Calculation}}. International Journal of Computer Vision 97(2):229--241. \doi{10.1007/s11263-011-0481-8}

\bibitem[{Viergever et~al.(2016)Viergever, Maintz, Klein, Murphy, Staring, and Pluim}]{viergever_survey_2016}
Viergever MA, Maintz JBA, Klein S, et~al (2016) A survey of medical image registration -- under review. Medical Image Analysis 33:140--144. \doi{10.1016/j.media.2016.06.030}

\bibitem[{Younes(2019)}]{younes_shapes_2019}
Younes L (2019) Shapes and {{Diffeomorphisms}}, Applied {{Mathematical Sciences}}, vol 171. Springer Berlin Heidelberg, Berlin, Heidelberg, \doi{10.1007/978-3-662-58496-5}

\end{thebibliography}
